\documentclass[]{siamart190516}

\usepackage{amsfonts}                                            % for \mathbb

\usepackage{bm}                                % for \bm, \bmdefine, bold math

\usepackage[caption=false]{subfig}

\usepackage{textcomp}   % for \textregistered (registered trademark) on MATLAB

\usepackage{exscale,relsize}                % for \mathlarger and \mathsmaller

\bmdefine{\ehat}{e}
\bmdefine{\nhat}{n}

\bmdefine{\nuhat}{\nu}

\bmdefine{\bmC}{C}
\bmdefine{\bmD}{D}
\bmdefine{\bmE}{E}
\bmdefine{\bmH}{H}
\bmdefine{\bmP}{P}

\bmdefine{\bmd}{d}
\bmdefine{\bmu}{u}
\bmdefine{\bmv}{v}

\bmdefine{\bmeps}{\epsilon}

\bmdefine{\bmchi}{\chi}

\newcommand{\calF}{\mathcal{F}}
\newcommand{\calN}{\mathcal{N}}
\newcommand{\calU}{\mathcal{U}}
\newcommand{\calV}{\mathcal{V}}

\newcommand{\calFE}{\calF_\text{E}}
\newcommand{\calFH}{\calF_\text{H}}

\newcommand{\Ftilde}{\widetilde{\calF}}

\newcommand{\chia}{\chi_\text{a}}

\newcommand{\chipara}{\chi_{\scriptscriptstyle\parallel}^{\text{e}}}
\newcommand{\chiperp}{\chi_{\scriptscriptstyle\perp}^{\text{e}}}

\newcommand{\eps}{\varepsilon}

\newcommand{\epsa}{\epsilon_\text{a}}
\newcommand{\epsz}{\epsilon_0}
\newcommand{\epara}{\epsilon_{\scriptscriptstyle\parallel}}
\newcommand{\eperp}{\epsilon_{\scriptscriptstyle\perp}}

\newcommand{\Ec}{E_\text{c}}
\newcommand{\Hc}{H_\text{c}}
\newcommand{\Hp}{H_\text{p}}

\newcommand{\ahat}{a}
\newcommand{\bhat}{b}
\newcommand{\chat}{c}
\newcommand{\dhat}{d}

\newcommand{\nhatb}{\nhat_\text{b}}
\newcommand{\nhats}{\nhat_\text{s}}

\newcommand{\We}{W_\text{e}}
\newcommand{\Ws}{W_\text{s}}
\newcommand{\WE}{W_\text{E}}
\newcommand{\WH}{W_\text{H}}

\newcommand{\bfI}{\mathbf{I}}
\newcommand{\bfP}{\mathbf{P}}

\newcommand{\bfu}{\mathbf{u}}
\newcommand{\bfx}{\mathbf{x}}

\newcommand{\bfzero}{\mathbf{0}}

\newcommand{\chitensor}{\bmchi^{\text{e}}}

\newcommand{\epstensor}{{\mathlarger\bmeps}}

\newcommand{\Rn}{\mathbb{R}^n}

\newcommand{\dWdn}{\frac{\partial W}{\partial\nhat}}
\newcommand{\dWedn}{\frac{\partial\We}{\partial\nhat}}
\newcommand{\dWsdn}{\frac{\partial\Ws}{\partial\nhat}}
\newcommand{\dWdgn}{\frac{\partial W}{\partial\nabla\nhat}}
\newcommand{\dWedgn}{\frac{\partial\We}{\partial\nabla\nhat}}

\newcommand{\PHI}{{\mathlarger\Phi}}
\newcommand{\PSI}{{\mathlarger\Psi}}

\newcommand{\Freed}{Fr\'{e}edericksz}

\newcommand{\Pf}{\bmP_\text{f}}

\newcommand*{\eb}{e_\text{b}}
\newcommand*{\es}{e_\text{s}}

\newlength{\ruleht}                                 % elevation of the overbar
\newlength{\rulewd}                                 % length of the overbar

\newcommand{\overbar}[4]{%
  \settoheight{\ruleht}{\ensuremath{#4}}%    init \ruleht to height of #4 box
  \addtolength{\ruleht}{#2}%                 add TOP SHIFT to \ruleht
  \settowidth{\rulewd}{\ensuremath{#4}}%     init \rulewd to width of #4 box
  \addtolength{\rulewd}{-#1}%                subtract LEFT  SHIFT from \rulewd
  \addtolength{\rulewd}{-#3}%                subtract RIGHT SHIFT from \rulewd
  \hspace{#1}%
  \raisebox{\ruleht}[0pt][0pt]{\makebox[0pt][l]{\rule{\rulewd}{.4pt}}}%
  \hspace{-#1}%
  \ensuremath{#4}}

\newcommand{\Hbar}{\overbar{1.75pt}{1pt}{.25pt}{H}}

\newcommand{\Ibar}{\overbar{1.5pt}{1pt}{-.25pt}{I}}

\newcommand{\Kbar}{\overbar{1.75pt}{1pt}{0pt}{K}}

\newcommand{\Lbar}{\overbar{1.5pt}{1pt}{.5pt}{L}}

\newcommand{\abar}{\overbar{1pt}{1pt}{.5pt}{a}}
\newcommand{\dbar}{\overbar{1.5pt}{.75pt}{0pt}{d}}
\newcommand{\qbar}{\overbar{.5pt}{1pt}{0pt}{q}}
\newcommand{\qbarsub}{\overbar{.5pt}{-.5pt}{.75pt}{q}}
\newcommand{\zbar}{\overbar{1.25pt}{1pt}{.25pt}{z}}

\newcommand{\lambdabar}{\overbar{.75pt}{1pt}{.25pt}{\lambda}}

\newcommand{\Omegabar}{\overbar{.75pt}{.75pt}{.75pt}{\Omega}}

\let\div\relax
\DeclareMathOperator{\div}{div}

\DeclareMathOperator{\curl}{curl}

\DeclareMathOperator{\tr}{tr}

\title{Electric-Field-Induced Instabilities \\
  in Nematic Liquid Crystals\thanks{Submitted to the editors \today}}

\author{Eugene C. Gartland, Jr.\thanks{Department of Mathematical
    Sciences, Kent State University, P.O.~Box 5190, Kent, OH 44242
    (\email{gartland@math.kent.edu},
    \url{http://www.math.kent.edu/\string~gartland}).}}

\headers{Electric-Field-Induced Instabilities in Liquid
  Crystals}{E. C. Gartland, Jr.}

\begin{document}

\maketitle

\begin{abstract}
  Systems involving nematic liquid crystals subjected to magnetic
  fields or electric fields are modeled using the Oseen-Frank
  macroscopic continuum theory, and general criteria are developed to
  assess the local stability of equilibrium solutions.  The criteria
  take into account the inhomogeneity of the electric field (assumed
  to arise from electrodes held at constant potential) and the mutual
  influence of the electric field and the liquid-crystal director
  field on each other.  The criteria show that formulas for the
  instability thresholds of electric-field \Freed\ transitions cannot
  in all cases be obtained from those for the analogous magnetic-field
  transitions by simply replacing the magnetic parameters by the
  corresponding electric parameters, contrary to claims in standard
  references.  This finding is consistent with observations made in
  [Arakelyan, Karayan, Chilingaryan, Sov.\ Phys.\ Dokl., \textbf{29}
  (1984), pp.\,202--204].  A simple analytical test is provided to
  determine when an electric-field-induced instability can differ
  qualitatively from the analogous magnetic-field-induced instability;
  the test depends only on the orientations of the ground-state fields
  and their admissible variations.  For the systems we study, it is
  found that taking into account the full coupling between the
  electric field and the director field can either elevate or leave
  unchanged an instability threshold (never lower it), compared to the
  threshold provided by the magnetic-field analogy (i.e., compared to
  treating the electric field as a uniform external force field).  The
  physical mechanism that underlies the effect of elevating an
  instability threshold is the added free energy associated with a
  first-order change in the ground-state electric field caused by a
  perturbation of the ground-state director field.  Examples are given
  that involve classical \Freed\ transitions and also periodic
  instabilities, with the periodic instability of Lonberg and Meyer
  [Phys.\ Rev.\ Lett., \textbf{55} (1985), pp.\,718--721] being
  further explored.  The inclusion of flexoelectric terms in the
  theory is studied, and it is found that these terms are not capable
  of altering the instability thresholds of any of the classical
  \Freed\ transitions, consistent with known results for the cases of
  the magnetic-field and the electric-field splay transitions.
\end{abstract}


% candidates for Subject Classification:
% 
%   49    calculus of variations and optimal control; optimization
%   49K     optimality conditions
%   49K20     problems involving PDEs
%   49K35     minimax problems
%   49K40     sensitivity, stability, well posedness
%   49S     variational principles of physics
%   49S05     variational principles of physics (should also be assigned
%             at least one other classification number in section 49)
%   76    fluid mechanics
%   76A     foundations, constitutive equations, rheology
%   76A15     liquid crystals
%   78    optics, electromagnetic theory
%   78A     general
%   78A30     electro- and magneto-statics
%   82    statistical mechanics, structure of matter
%   82B     equilibrium statistical mechanics
%   82B21     continuum models (systems of particles, etc.)
%   82D     applications to specific types of physical systems
%   82D30     random media, disordered materials (including liquid
%             crystals and spin glasses)


\section{Introduction}

%\label{sec:introduction}

Our interest is in macroscopic continuum models for the orientational
properties of materials in a liquid crystal phase, a complex partially
ordered fluid phase exhibited by certain materials in certain
parameter ranges.  Such models are used at the scales of typical
devices and experiments involving these kinds of materials
(micrometer-scale thin films, and the like).  Liquid crystals are very
responsive to external stimuli, such as magnetic fields and electric
fields, and this has been one of the keys to their usefulness in
technological applications.  This response frequently manifests itself
in an instability such that an abrupt change in the orientational
properties of the material occurs at a critical threshold of the
strength of the applied magnetic or electric field, the textbook
examples of this being ``\Freed\ transitions''---see
\cite[\S3.2.3]{degennes:prost:93} or \cite[\S3.4]{stewart:04} or
\cite[\S4.2]{virga:94}.  Our main objective is to illuminate
differences between instabilities induced by magnetic fields versus
those induced by electric fields.  We do this via the development of
stability criteria that take into account the inhomogeneity of the
electric field and its coupling to the orientational properties of the
material.  The characterizations of local stability that we develop
mimic familiar results found in equality-constrained optimization
theory in $\Rn$.

At the macroscopic level of modeling, the orientational state of a
material in a uniaxial nematic liquid crystal phase is characterized
by a unit-length vector field $\nhat$ (the ``director field''), which
represents the average orientation of the distinguished axis of the
anisometric molecules in a fluid element at a point.  Central to the
modeling of equilibrium configurations of the director field is an
appropriate expression for the free energy of the system, a
thermodynamic potential that serves as a work function for isothermal,
reversible processes.  In the models of interest to us, the material
is assumed to be incompressible.  For simplicity, we restrict our
attention to achiral uniaxial nematic liquid crystals (the simplest
liquid crystal phase).  Such materials are characterized by
intermolecular forces that encourage parallel alignment of directors,
leading to uniform ground-state director fields.  Other influences
(boundary conditions, external force fields) can effect nonuniform
equilibrium configurations of $\nhat$, at a cost of distortional
elastic energy.  Details are presented in what follows.  Standard
references include \cite{degennes:prost:93,stewart:04,virga:94}.

The force fields of external origin most commonly encountered in the
context of liquid crystals are magnetic fields and electric fields.
Magnetic fields are influenced by a liquid crystal medium, which is
anisotropic with magnetic susceptibilities that depend on the
orientational state of the material at a point.  For the parameter
values of typical liquid crystal materials, however, this influence is
negligible \cite{arakelyan:karayan:chilingaryan:84},
\cite[\S2.1]{gartland:20}.  Thus, a magnetic field in a liquid crystal
can be treated as a uniform external field.  An electric field is
influenced by the state of the liquid crystal material in a similar
fashion; however, the coupling is much stronger and should not be
ignored \cite{arakelyan:karayan:chilingaryan:84},
\cite[\S2.1]{gartland:20}.  Thus, the equilibrium state of a liquid
crystal subjected to an electric field should be determined in a
self-consistent way, with the director field and the electric field
treated as coupled state variables.  This coupling in general leads to
inhomogeneities of the electric field and complicates the
determination of equilibrium fields and the assessment of their local
stability properties.

While the differences between magnetic fields and electric fields in
liquid crystals have been appreciated for some time, the widely held
view is that they give rise to only modest quantitative differences
but not to qualitative differences in the context of instabilities
such as \Freed\ transitions.  For example, in
\cite[\S3.3.1]{degennes:prost:93} (referencing \cite{gruler:meier:72})
and \cite[\S3.5]{stewart:04}, it is asserted that electric-field
\Freed\ thresholds can be obtained from the formulas for
magnetic-field thresholds by simply substituting the electric
parameters for the corresponding magnetic parameters.  In fact, this
was borne out in \cite{deuling:72}, where the electric-field \Freed\
transition in a particular geometry was analyzed taking into account
the full coupling between the director field and the electric field.
There it was found that in contrast to the approximation by a
\emph{uniform} electric field, slightly smaller values were obtained
for the distortion of the liquid crystal director field past the onset
of the instability, though the critical threshold of the electric
field itself was the same in the coupled case as in the case of the
approximation by a uniform external electric field (consistent with
the recipe of \cite[\S3.3.1]{degennes:prost:93} and
\cite[\S3.5]{stewart:04}).  The analysis of \cite{deuling:72} is
recounted in \cite[\S3.5]{stewart:04}.

A common use of the threshold formulas for the various \Freed\
transitions is in determining via experimental measurements certain
material-dependent parameters of different liquid crystals
\cite[\S3.2.3.1]{degennes:prost:93}.  Such experiments can be done
with magnetic fields or with electric fields, whichever is more
convenient, and experimentalists invariably take for granted the
validity of the simple relationships between the formulas for the
magnetic-field threshold versus the electric-field threshold implied
by the recipes of \cite[\S3.3.1]{degennes:prost:93} and
\cite[\S3.5]{stewart:04}---see, for example,
\cite{bradshaw:raynes:bunning:faber:85} or
\cite[Ch.\,5]{dunmur:fukuda:luckhurst:01} for more discussion and
additional references.  Here we will show that true qualitative
differences can occur between magnetic-field-induced instabilities and
those induced by electric fields (such as instability thresholds that
differ from the recipes of \cite[\S3.3.1]{degennes:prost:93} and
\cite[\S3.5]{stewart:04}), and we provide simple criteria to identify
them.  We also provide illustrative examples.  Our results expand upon
ideas in \cite{arakelyan:karayan:chilingaryan:84} and
\cite{rosso:virga:kralj:04}.

The paper is organized as follows.  In \cref{sec:model-problems}, we
introduce models involving magnetic fields and models involving
electric fields (free energies, domains, boundary conditions).
Stability criteria for the magnetic-field models are developed in
\cref{sec:stability-for-H-fields}.  These take the form of first-order
and second-order necessary conditions, in the spirit of
equality-constrained optimization theory in $\Rn$.  Illustrations are
given involving classical \Freed\ transitions as well as periodic
instabilities.  \Cref{sec:stability-for-E-fields} extends these
results to the models involving electric fields, which introduces new
aspects.  There, examples are given illustrating the types of
qualitative differences that occur in certain classical instabilities
when induced by electric fields as opposed to magnetic fields.  In
\cref{sec:conclusions}, we summarize our main results.
\Cref{app:lonberg-meyer} contains details of the analysis of one of
the examples involving a periodic instability (that of Lonberg and
Meyer \cite{lonberg:meyer:85}), and \cref{app:flexoelectric} provides
an illustration of how the approach can be extended to include an
additional feature (flexoelectric effects) in the model.  It is also
shown in \cref{app:flexoelectric} that the additional flexoelectric
terms in the free energy do not influence the instability thresholds
of any of the classical \Freed\ transitions.

\section{Model problems}

\label{sec:model-problems}

\begin{figure}
  \centering
  \includegraphics[width=.75\linewidth]{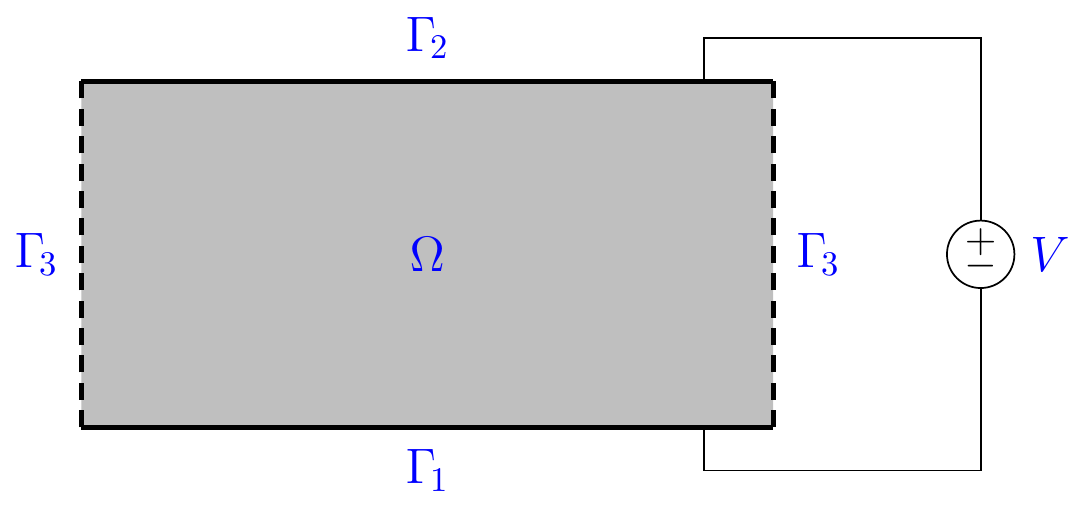}
  \caption{Model problem domain (two-dimensional depiction of actual
    three-dimensional, hexahedral domain).  Boundary conditions: on
    $\Gamma_1$ (strong anchoring, Dirichlet boundary condition
    $\nhat=\nhatb$ on $\nhat$, $\varphi=0$), on $\Gamma_2$ (weak
    anchoring, natural boundary condition $(\ref{eqn:strong})_2$ on
    $\nhat$, $\varphi=V$), on $\Gamma_3$ (periodic boundary conditions
    on both $\nhat$ and $\varphi$ on opposing sides of $\Gamma_3$).}
  \label{fig:domain}
\end{figure}

We perform our analyses on two model problems, one for a system with a
magnetic field, the other for one with an electric field.  Both
problems share the same domain and boundary conditions on the director
field $\nhat$.  The domain $\Omega$ is hexahedral (as shown in
\cref{fig:domain}),
with lower boundary $\Gamma_1$, upper boundary $\Gamma_2$, and lateral
boundary $\Gamma_3$.  A ``strong anchoring condition'' (Dirichlet
boundary condition) is imposed on $\nhat$ on $\Gamma_1$, a ``weak
anchoring condition'' (a natural boundary condition resulting from a
surface anchoring energy) on $\Gamma_2$, and periodic boundary
conditions on opposing sides of $\Gamma_3$.  These cover the three
types of boundary conditions typically encountered in modeling liquid
crystal systems.  The model domain may be viewed as a subdomain in a
liquid crystal thin film regarded as having infinite extent in the
lateral directions.% The models we employ here
%are adapted from \cite[\S2]{gartland:20}.

For a uniaxial nematic liquid crystal subject to a magnetic field
(with boundary conditions as described above), the standard model free
energy is an integral functional of the director field that can be
taken in the form
\begin{equation}\label{eqn:Fn}
  \calF[\nhat] =
  \int_\Omega W(\nhat,\nabla\nhat) \, dV +
  \int_{\Gamma_2} \Ws(\nhat) \, dS ,
\end{equation}
where $W$ denotes the free-energy density (per unit volume) and $\Ws$
the surface-anchoring energy (per unit area).  The free-energy density
consists of a part associated with distortional elasticity $\We$ (for
which we employ the classical Oseen-Frank formula) and a part
associated with the magnetic induction, $\WH$:
\begin{equation*}
  W = \We(\nhat,\nabla\nhat) + \WH(\nhat) ,
\end{equation*}
with
\begin{equation}\label{eqn:We}
  2\We =
  K_1(\div\nhat)^2 + K_2(\nhat\cdot\curl\nhat)^2 +
  K_3|\nhat\times\curl\nhat|^2 +
  K_{24} \bigl[ \tr(\nabla\nhat)^2 - (\div\nhat)^2 \bigr] .
\end{equation}
Here $K_1$, $K_2$, $K_3$, and $K_{24}$ are material-dependent
parameters (``elastic constants''), which under appropriate conditions
($0 < K_1, K_2, K_3$ and $0 < K_{24} < 2 \min\{K_1,K_2\}$) guarantee
that $\We\ge0$ and $\We=0$ if and only if $\nabla\nhat=\bfzero$.  The
constants $K_1$, $K_2$, and $K_3$ are referred to as the ``splay,''
``twist,'' and ``bend'' constants, because of the simple types of
distortions that they penalize (see \cite[\S3.1.1]{degennes:prost:93}
or \cite[\S2.2]{stewart:04} or \cite[\S3.3]{virga:94}); this
terminology will come up in some of our examples.  The $K_{24}$ term
is a null Lagrangian and does not play a role in many simple systems.
The precise form of $\We$ does not matter to our development (though
terms from it appear in examples that will follow).  The simplest form
for $\We$ corresponds to $K_1=K_2=K_3=K_{24}=K$, which gives
$\We = \frac12 K |\nabla\nhat|^2$ (the ``equal elastic constants
model'').

The contribution to the free energy associated with the magnetic field
(for materials of the type we study here) can be taken in the form
\begin{equation*}
  \WH = - \frac12 \chia ( \bmH \cdot \nhat )^2 ,
\end{equation*}
with $\chia$ the diamagnetic anisotropy (a material-dependent
parameter that can be positive or negative) and $\bmH$ the magnetic
field (which can be assumed to be constant, as discussed in the
Introduction).
% The parameter $\chia$ is sometimes denoted
% $\mu_0\Delta\chi$, with $\mu_0$ the free-space magnetic permeability.
Globally stable configurations of the director field minimize $\calF$
(subject to boundary conditions and the pointwise constraint
$|\nhat|=1$); so $\chia>0$ encourages the director to be parallel to
$\bmH$, while $\chia<0$ encourages it to be perpendicular to $\bmH$.
The surface anchoring energy $\Ws$ can take a variety of forms, a
simple example being
\begin{equation*}
  \Ws = - \frac12 W_0 ( \nhat \cdot \nhats )^2 ,
\end{equation*}
with $W_0$ the ``anchoring strength.''  With $W_0>0$, this encourages
$\nhat$ to be parallel to the prescribed orientation $\nhats$ (the
``easy axis'') on the boundary.  See \cite[\S2.2,
App.\,A.1]{gartland:20} for more examples and references.  The
modeling aspects above are well documented in
\cite[\S\S3.1,\,3.2]{degennes:prost:93},
\cite[\S\S2.2,\,2.3,\,2.6]{stewart:04}, and
\cite[\S\S3.2,\,3.5,\,4.1]{virga:94}.  A summary of the relevant
points is in \cite[\S2]{gartland:20}, from which we have adapted our
model problems.

For a nematic liquid crystal subject to an electric field, the mutual
influence of the electric field on the director field and of the
director field on the electric field should be taken into account, as
discussed in the Introduction.  We do this by employing a free energy
that is a functional of two state variables: the director field
$\nhat$ and the electric potential field $\varphi$ (related to the
electric field via $\bmE = - \nabla\varphi$).  The free energy now
takes the form
\begin{equation*}
  \calF[\nhat,\varphi] =
  \int_\Omega W(\nhat,\nabla\nhat,\nabla\varphi) \, dV +
  \int_{\Gamma_2} \Ws(\nhat) \, dS ,
\end{equation*}
with
\begin{equation*}
  W = \We(\nhat,\nabla\nhat) + \WE(\nhat,\nabla\varphi) .
\end{equation*}
Here $\Omega$, $\Gamma_2$, $\We$, and $\Ws$ are exactly as before, and
the relevant relations from electrostatics are given by
% \begin{gather*}
%   \WE = - \frac12 \bmD \cdot \bmE , \quad
%   \bmD = \epstensor(\nhat) \bmE , \quad \bmE = - \nabla\varphi \\
%   \epstensor = \epsz \bigl[ \epara \bfI +
%   \epsa ( \nhat\otimes\nhat ) \bigr] , \quad \epsa := \epara - \eperp .
% \end{gather*}
\begin{equation}\label{eqn:WE}
  \WE = - \frac12 \bmD \cdot \bmE , \quad
  \bmD = \epstensor(\nhat) \bmE , \quad
  \epstensor = \epsz ( \eperp \bfI + \epsa \nhat\otimes\nhat ) , \quad
  \epsa := \epara - \eperp .
\end{equation}
Here $\bmD$ is the electric displacement, ${\mathlarger\epstensor}$
the dielectric tensor, $\epsz$ the vacuum dielectric constant, and
$\epara$ and $\eperp$ the material-dependent relative permittivities
parallel to and perpendicular to the local director.  The dielectric
anisotropy $\epsa$ can be positive or negative.
% (and is sometimes denoted $\Delta\eps$).
The expression for $\WE$ is the correct electric contribution to the
free energy associated with an electric field generated by electrodes
held at constant potential in a transversely isotropic linear
dielectric that contains no distribution of free charge.  The electric
potential $\varphi$ satisfies Dirichlet boundary conditions on
$\Gamma_1$ and $\Gamma_2$ (of prescribed difference $V$) and periodic
boundary conditions on opposing sides of $\Gamma_3$.  For more
discussion and additional references, see
\cite[\S3.3]{degennes:prost:93}, \cite[\S2.3.1]{stewart:04}, and
\cite[\S4.1]{virga:94}, or the synopsis in \cite[\S2.1]{gartland:20}.

To summarize, our two model free energies are
%\begin{subequations}
\begin{gather}
  \calFH[\nhat] = \int_\Omega \Bigl[ \We(\nhat,\nabla\nhat) -
  \frac12 \chia (\bmH\cdot\nhat)^2 \Bigr] dV +
  \int_{\Gamma_2} \Ws(\nhat) \, dS \label{eqn:FH} \\
  \calFE[\nhat,\varphi] = \int_\Omega \Bigl[ \We(\nhat,\nabla\nhat) -
  \frac12 \epstensor(\nhat) \nabla\varphi \cdot \nabla\varphi \Bigr] dV +
  \int_{\Gamma_2} \Ws(\nhat) \, dS , \label{eqn:FE}
\end{gather}
%\end{subequations}
with $\Omega$ and $\Gamma_2$ as depicted in \cref{fig:domain}, $\We$
as in \cref{eqn:We}, $\Ws$ an appropriate surface anchoring energy,
and ${\mathlarger\epstensor}$ as in \cref{eqn:WE}.  Equilibrium fields
are stationary points of these functionals (subject to the essential
boundary conditions and the pointwise constraint $|\nhat|=1$), with
globally stable phases corresponding to equilibrium fields of least
free energy.  The characterization of local stability of equilibria is
the main topic that we take up in what follows.  We note that since
the dielectric tensor ${\mathlarger\epstensor}$ is symmetric positive
definite, the stationary points of $\calFE$ are maximizing with
respect to $\varphi$, though they are locally minimizing with respect
to $\nhat$.

In what follows, we assume that all admissible fields and admissible
variations are regular enough to satisfy the various equilibrium
characterizations in strong forms.  We do this for simplicity and note
that it precludes the presence of any singularities (``defects'' or
``disclinations'') in the systems we study.  The models that we deal
with are vectorial in nature with pointwise constraints and associated
Lagrange multiplier fields of low regularity in the presence of
defects.  Weak variational formulations can require cumbersome
technical assumptions---see, for example,
\cite{hu:tai:winther:09,ito:kunish:08}.  Let $C^2(\Omegabar)$ denote
real-valued scalar fields on $\Omega$ that are twice continuously
differentiable with finite limits on $\partial\Omega$ (of the fields
and their derivatives up to second order), and let $\bmC^2(\Omegabar)$
denote the analogous space for vector fields on $\Omega$ (with values
in real three-dimensional Euclidean space).  Such fields are more than
smooth enough for our purposes and are sufficient to ensure that the
Lagrange multiplier fields we encounter will be bounded and
continuous.  We define our classes of admissible fields for $\nhat$
and $\varphi$ as follows:
\begin{gather*}
  \calN =
  \bigl\{ \nhat \in \bmC^2(\Omegabar) \, \bigl| \,
  |\nhat| = 1 \text{ in } \Omega, \nhat = \nhatb \text{ on } \Gamma_1 ,
  \nhat \text{ periodic on } \Gamma_3 \bigr\} \\
  \PHI =
  \bigl\{ \varphi \in C^2(\Omegabar) \, \bigl| \,
  \varphi = 0 \text{ on } \Gamma_1 , \varphi = V \text{ on } \Gamma_2 ,
  \varphi \text{ periodic on } \Gamma_3 \bigr\} .
\end{gather*}
Periodic here is taken to mean periodic on opposing sides of the
lateral boundary of the hexahedral domain.

With our notation now defined, we can succinctly characterize globally
stable solutions of our two models problems as follows:
\begin{equation*}
  \calFH[\nhat^*] = \min_{\nhat\in\calN} \calFH[\nhat] , \quad
  \calFE[\nhat^*\!\!,\varphi^*] =
  \min_{\nhat\in\calN} \max_{\varphi\in\Phi} \calFH[\nhat,\varphi] .
\end{equation*}
Lacking convexity, these systems can have more than one globally
stable solution.  While the electric-field problems have an intrinsic
minimax nature, their globally stable solutions still admit a
characterization by a ``least free energy principle'': a globally
stable solution pair $\nhat^*$, $\varphi^*$ is an equilibrium pair of
least free energy (among all equilibrium pairs).  This point of view
will be found to be useful in what follows.

\section{Stability criteria for magnetic fields}

\label{sec:stability-for-H-fields}

If a liquid crystal system is sufficiently simple, then the local
stability of an equilibrium director configuration of $\calFH$ can be
analyzed by representing the director field in terms of one or two
orientation angles (e.g.,
$\nhat = \cos\theta\,\ehat_1 + \sin\theta\,\ehat_2$).  Such
representations free one from having to deal with the pointwise
constraint $|\nhat|=1$ (which presents a complicating factor for
numerical modeling, as well as for analysis).  With the free energy
expressed in terms of orientation angles, local stability is simply
assessed in terms of the positive definiteness of the second variation
of the free-energy functional.  If, on the other hand, one chooses to
(or needs to) model the director in terms of its components with
respect some frame, then one must enforce $|\nhat|=1$ pointwise, and
the Lagrange multiplier field associated with this enters both the
equilibrium Euler-Lagrange equations as well as the criteria for local
stability, as observed in \cite{rosso:virga:kralj:04} and as we shall
see below.

Analyses using director components have been used in the past to study
the stability of specific configurations, such as radial point defects
(``hedgehogs'')---see, for example,
\cite{cohen:luskin:91,cohen:taylor:90,helein:87,kinderlehrer:ou:92}.
A stability criterion of a general nature is presented in
\cite{rosso:virga:kralj:04}, and it is closely related to our results
for the case of a magnetic field (though here it is somewhat
differently expressed and derived).  A main contribution here is the
extension of such ideas to systems involving coupled electric fields.
Minimizing $\calFH$ subject to $|\nhat|=1$ can be viewed as the
continuum analogue of a problem in equality-constrained optimization
in $\Rn$, and we pursue this analogy, beginning with a recapitulation
of the relevant formulas from that area.

\subsection{Results from equality-constrained optimization theory}

A discrete analogue of the constrained minimization problem for
$\calFH$ is provided by the following:
\begin{equation*}
  \min_{\bfx\in\Rn} f(\bfx) , \quad
  \text{subject to } g_1(\bfx) = \cdots = g_m(\bfx) = 0 .
\end{equation*}
Here the objective function $f$ and constraint functions
$g_1,\ldots,g_m$ are assumed to be real valued and smooth, and $\Rn$
denotes real $n$-dimensional Euclidean space with the standard inner
product.  The first-order and second-order necessary conditions
associated with a local solution $\bfx_0$ are as follows.  Under mild
non-degeneracy conditions (such as linear independence of $\nabla
g_1(\bfx_0),\ldots,\nabla g_m(\bfx_0)$), there exist unique Lagrange
multipliers $\lambda_1^0,\ldots,\lambda_m^0\in\mathbb{R}$ such that
\begin{equation}\label{eqn:discrete-first-order}
  \nabla f(\bfx_0) = \lambda_1^0 \nabla g_1(\bfx_0) + \cdots +
  \lambda_m^0 \nabla g_m(\bfx_0)
\end{equation}
and
\begin{subequations}\label{eqn:discrete-second-order}
  \begin{equation}\label{eqn:discrete-second-order-a}
    \bigl[ \nabla^2 f(\bfx_0) - \lambda_1^0 \nabla^2 g_1(\bfx_0) -
    \cdots - \lambda_m^0 \nabla^2 g_m(\bfx_0) \bigr]
    \bfu \cdot \bfu \ge 0 ,
  \end{equation}
  for all $\bfu \in \Rn$ satisfying
  \begin{equation}\label{eqn:discrete-second-order-b}
    \nabla g_1(\bfx_0) \cdot \bfu = \cdots =
    \nabla g_m(\bfx_0) \cdot \bfu = 0 .
  \end{equation}
\end{subequations}
That is to say, the constrained stationary point will be a local
minimum only if the Hessian of the Lagrangian is positive semi
definite on the tangent space to the constraint manifold at the point.
Strict positivity in \cref{eqn:discrete-second-order-a} for nontrivial
$\bfu$ satisfying \cref{eqn:discrete-second-order-b} is sufficient for
local stability.  We sketch below an approach to deriving these
results that generalizes to our free-energy-minimization problems.
The results can be found in standard references on optimization
theory, such as
\cite{fletcher:87,gill:murray:wright:81,luenberger:69,marlow:78}.

The conditions above can be deduced as follows.  Give a trajectory
$\bfx(t)$ on the constraint manifold through $\bfx_0$ smoothly
parametrized by $t$:
\begin{equation*}
  g_1(\bfx(t)) = \cdots = g_m(\bfx(t)) = 0 , \quad
  - c < t < c , \text{ some } c > 0 , \quad
  \bfx(0) = \bfx_0 .
\end{equation*}
With the definition $F(t) := f(\bfx(t))$, the point $\bfx_0$ being a
local minimum implies $F'(0)=0$ and $F''(0)\ge0$.  Now
\begin{equation*}
  F'(0) = \nabla f(\bfx_0) \cdot \dot{\bfx}_0 ~~ \text{and} ~~
  F''(0) = \nabla f(\bfx_0) \cdot \ddot{\bfx}_0 +
  \nabla^2 f(\bfx_0) \dot{\bfx}_0 \cdot \dot{\bfx}_0 ,
\end{equation*}
where
\begin{equation*}
  \dot{\bfx}_0 := \frac{d}{dt} \bfx(t) \bigr|_{t=0} ~~ \text{and} ~~
  \ddot{\bfx}_0 := \frac{d^2}{dt^2} \bfx(t) \bigr|_{t=0} .
\end{equation*}
For each of the constraints, we have
\begin{equation*}
  g_i(\bfx(t)) = 0 ~ \Rightarrow ~
  \frac{d}{dt} g_i(\bfx(t)) = \frac{d^2}{dt^2} g_i(\bfx(t))
  = \cdots = 0 , ~ - c < t < c ,
\end{equation*}
from which we obtain,
\begin{equation*}
  \nabla g_i(\bfx_0) \cdot \dot{\bfx}_0 = 0 , ~~
  \nabla g_i(\bfx_0) \cdot \ddot{\bfx}_0 +
  \nabla^2 g_i(\bfx_0) \dot{\bfx}_0 \cdot \dot{\bfx}_0 = 0 , ~~
  \text{for } i = 1, \ldots, m .
\end{equation*}
Thus $\dot{\bfx}_0$ is in the tangent space to the constraint manifold
at $\bfx_0$, and stationarity implies
$F'(0) = \nabla f(\bfx_0) \cdot \dot{\bfx}_0 = 0$, for all such
$\dot{\bfx}_0$ as well.  Assuming the constraint normals
$\nabla g_1(\bfx_0),\ldots,\nabla g_m(\bfx_0)$ to be linearly
independent, for example, this guarantees that $\nabla f(\bfx_0)$ has
a unique representation as a linear combination of
$\nabla g_1(\bfx_0), \ldots, \nabla g_m(\bfx_0)$, i.e.,
\cref{eqn:discrete-first-order} holds with unique
$\lambda_1^0,\ldots,\lambda_m^0$.  This relation and the second part
of the relations above can be used to simplify the requirement
\begin{equation*}
  0 \le F''(0) = \nabla f(\bfx_0) \cdot \ddot{\bfx}_0 +
  \nabla^2 f(\bfx_0) \dot{\bfx}_0 \cdot \dot{\bfx}_0
\end{equation*}
via
\begin{align*}
  \nabla f(\bfx_0) \cdot \ddot{\bfx}_0 &=
  \lambda_1^0 \nabla g_1(\bfx_0) \cdot \ddot{\bfx}_0 + \cdots +
  \lambda_m^0 \nabla g_m(\bfx_0) \cdot \ddot{\bfx}_0 \\
  &= - \lambda_1^0 \nabla^2 g_1(\bfx_0) \dot{\bfx}_0 \cdot \dot{\bfx}_0
     - \cdots -
       \lambda_m^0 \nabla^2 g_m(\bfx_0) \dot{\bfx}_0 \cdot \dot{\bfx}_0 .
\end{align*}
Substituting this into the inequality on $F''(0)$ above leads to the
second-order necessary condition \cref{eqn:discrete-second-order}.
One anticipates that it should be possible to frame the statements and
analysis of our continuum liquid crystal models in a similar way, and
we endeavor to do so below.

\subsection{Stability criteria}

We seek to establish similar conditions for a local minimum $\nhat_0$
of a functional of the form \cref{eqn:Fn},
\begin{equation*}
  \calF[\nhat] = \int_\Omega W(\nhat,\nabla\nhat) \, dV +
  \int_{\Gamma_2} \Ws(\nhat) \, dS ,
\end{equation*}
with $\Omega$ and $\Gamma_2$ as in \cref{fig:domain}, $W$ an
appropriate free-energy density, and $\Ws$ an appropriate anchoring
energy, subject to the essential boundary conditions of our model
problems (Dirichlet on $\Gamma_1$, periodic on $\Gamma_3$), and
subject to the pointwise constraint $|\nhat|=1$.  This includes the
model free energy $\calFH$ in \cref{eqn:FH}.  Let $\nhat_\eps$ be a
family of unit-length vector fields on $\Omega$ smoothly parametrized
by $\eps$ such that
\begin{equation*}
  |\nhat_\eps| = 1 , ~~
  - c < \eps < c , \text{ some } c > 0 , ~~
  \nhat_\eps |_{\eps=0} = \nhat_0 .
\end{equation*}
The most commonly used realization of such a field is
\begin{equation}\label{eqn:neps}
  \nhat_\eps = \frac{\nhat_0+\eps\bmv}{|\nhat_0+\eps\bmv|} ,
\end{equation}
where $\bmv$ is such that the combination $\nhat_0+\eps\bmv$ satisfies
the same essential boundary conditions and regularity assumptions that
$\nhat_0$ must satisfy but is otherwise arbitrary.  With the
definition $F(\eps) := \calF[\nhat_\eps]$, the point $\nhat_0$ being a
local minimum point implies $F'(0)=0$ and $F''(0)\ge0$.  Here
\begin{equation*}
  F'(0) = \delta \calF[\nhat_0](\dot{\nhat}_0) ~~ \text{and} ~~
  F''(0) = \delta \calF[\nhat_0](\ddot{\nhat}_0) +
  \delta^2 \! \calF[\nhat_0](\dot{\nhat}_0) ,
\end{equation*}
with $\delta\calF$ and $\delta^2\!\calF$ the first and second
variations,
\begin{equation*}
  \delta \calF[\nhat](\bmv) =
  \frac{d}{d\eps} \calF[\nhat+\eps\bmv] \bigr|_{\eps=0} ~~ \text{and} ~~
  \delta^2 \! \calF[\nhat](\bmv) =
  \frac{d^2}{d\eps^2} \calF[\nhat+\eps\bmv] \bigr|_{\eps=0} ,
\end{equation*}
giving
\begin{align*}
  \delta \calF[\nhat](\bmv) &= \int_\Omega \Bigl(
  \frac{\partial W}{\partial\nhat} \cdot \bmv +
  \frac{\partial W}{\partial\nabla\nhat} \cdot \nabla \bmv \Bigr) \, dV +
  \int_{\Gamma_2} \Bigl(
  \frac{\partial\Ws}{\partial\nhat} \cdot \bmv \Bigr) \, dS \\
  &= \int_\Omega \Bigl( \frac{\partial W}{\partial n_i} v_i +
  \frac{\partial W}{\,\partial n_{i,j}} v_{i,j} \Bigr) \, dV +
  \int_{\Gamma_2} \Bigl( \frac{\partial\Ws}{\partial n_i} v_i \Bigr) \, dS
\end{align*}
and
\begin{multline*}
  \delta^2 \! \calF[\nhat](\bmv) = \int_\Omega \Bigl(
  \frac{\partial^2W}{\partial n_i \partial n_k} v_i v_k +
  2 \frac{\partial^2W}{\partial n_i \partial n_{k,l}} v_i v_{k,l} +
  \frac{\partial^2W}{\partial n_{i,j} \partial n_{k,l}} v_{i,j} v_{k,l}
  \Bigr) \, dV \\
  {} + \int_{\Gamma_2} \Bigl(
  \frac{\partial^2W}{\partial n_i \partial n_k} v_i v_k \Bigr) \, dS ,
\end{multline*}
where
\begin{equation*}
  \dot{\nhat}_0 := \frac{d}{d\eps} \nhat_\eps \bigr|_{\eps=0}
  ~~ \text{and} ~~
  \ddot{\nhat}_0 := \frac{d^2}{d\eps^2} \nhat_\eps \bigr|_{\eps=0} .
\end{equation*}
Here $n_i$ and $v_i$ are the components of $\nhat$ and $\bmv$ with
respect to a fixed Cartesian frame,
$n_{i,j}=\partial n_i/\partial x_j$, etc., and summation over repeated
indices is implied.  We note that with $\nhat_\eps$ defined as in
\cref{eqn:neps}, we would have
\begin{equation*}
  \dot{\nhat}_0 = \bfP(\nhat_0) \bmv , \quad
  \bfP(\nhat) := \bfI - \nhat \otimes \nhat .
\end{equation*}
Given a unit-length vector field $\nhat$, the tensor field
$\bfP(\nhat)$ defined above projects transverse to $\nhat$ at each
point \cite[\S2.5]{virga:94} and is a convenient operator in the
analysis of director models.

Differentiations with respect to $\eps$ of $|\nhat_\eps|^2 =
\nhat_\eps \cdot \nhat_\eps = 1$ give rise to the pointwise relations
\begin{equation*}
  \nhat_0 \cdot \dot{\nhat}_0 = 0 , \quad
  \nhat_0 \cdot \ddot{\nhat}_0 + | \dot{\nhat}_0 |^2 = 0 .
\end{equation*}
Any such $\dot{\nhat}_0$ necessarily vanishes on $\Gamma_1$, is
periodic on opposing sides of $\Gamma_3$, and is transverse to
$\nhat_0$ (in the sense that $\nhat_0 \cdot \dot{\nhat}_0 = 0$ on
$\Omega$).  We denote the collection of all such vector fields
\begin{equation*}
  \calU_0 = \bigl\{ \bmu \in \bmC^2(\Omegabar) \, \bigl| \,
  \bmu = \bfzero \text{ on } \Gamma_1 ,
  \bmu \text{ periodic on } \Gamma_3 ,
  \nhat_0 \cdot \bmu = 0 \text{ on } \Omega \bigr\} .
\end{equation*}
Such vector fields can be generated from the larger class
\begin{equation*}
  \calV_0 = \bigl\{ \bmv \in \bmC^2(\Omegabar) \, \bigl| \,
  \bmv = \bfzero \text{ on } \Gamma_1 ,
  \bmv \text{ periodic on } \Gamma_3 \bigr\}
\end{equation*}
by using the transverse projector $\bfP(\nhat_0)$ above:
% \begin{equation*}
%   \bmv \in \calV_0 , ~~ \bmu = \bfP(\nhat_0) \bmv ,
%   ~~ \Rightarrow ~~ \bmu \in \calU_0 .
% \end{equation*}
\begin{equation*}
  \bmu \in \calU_0 ~ \Leftrightarrow ~
  \bmu = \bfP(\nhat_0) \bmv , \text{ some } \bmv \in \calV_0 .
\end{equation*}

The first-order necessary conditions follow from
\begin{equation*}
  F'(0) = 0 ~ \Rightarrow ~
  \delta \calF[\nhat_0](\bmu) = 0 , ~ \forall \bmu \in \calU_0 ,
\end{equation*}
which (using $\bmu = \bfP(\nhat_0) \bmv$ and integration by parts) can
be written in the following equivalent forms:
\begin{equation}\label{eqn:first-order}
  \delta \calF[\nhat_0](\bmv) =
  \int_\Omega \lambda_0 \nhat_0 \cdot \bmv \, dV +
  \int_{\Gamma_2} \mu_0 \nhat_0 \cdot \bmv \, dS , ~~
  \forall \bmv \in \calV_0
\end{equation}
or
\begin{equation}\label{eqn:strong}
  - \div \Bigl( \dWdgn \Bigr) + \dWdn =
  \lambda_0 \nhat_0  \text{ in } \Omega , \quad
  \Bigl( \dWdgn \Bigr) \nuhat + \dWsdn =
  \mu_0 \nhat_0 \text{ on } \Gamma_2 .
\end{equation}
\Cref{eqn:first-order} is the analogue of
\cref{eqn:discrete-first-order}.  The role of the constraint functions
$g_i$ and their gradients is here played by
\begin{equation*}
  g(\nhat) = \frac12 \bigl( |\nhat|^2 - 1 \bigr) , \text{ for which }
  \frac{\partial g}{\partial\nhat} = \nhat , ~
  \frac{\partial^2 g}{\,\partial\nhat^2} = \bfI .
\end{equation*}
The Lagrange multiplier fields $\lambda_0$ and $\mu_0$ are given by
\begin{equation*}
  \lambda_0 = \Bigl[
  - \div \Bigl( \dWdgn \Bigr) + \dWdn \Bigr] \cdot \nhat_0 , \quad
  \mu_0 = \Bigl[
  \Bigl( \dWdgn \Bigr) \nuhat + \dWsdn \Bigr] \cdot \nhat_0 ,
\end{equation*}
with the bracketed expressions evaluated on the equilibrium field
$\nhat_0$.  The results above are well known; the only point here is
to highlight the analogy to the discrete setting and to anticipate the
next steps.

The second-order necessary conditions follow from
\begin{equation*}
  F''(0) \ge 0 ~~ \Rightarrow ~~
  \delta \calF[\nhat_0](\ddot{\nhat}_0) +
  \delta^2 \calF[\nhat_0](\dot{\nhat}_0) \ge 0 .
\end{equation*}
The weak-form Euler-Lagrange equation \cref{eqn:first-order} and the
pointwise relation $\nhat_0\cdot\ddot{\nhat}_0+|\dot{\nhat}_0|^2=0$
can be used to simplify this as follows,
\begin{align*}
  \delta \calF[\nhat_0](\ddot{\nhat}_0) &=
  \int_\Omega \lambda_0 \nhat_0 \cdot \ddot{\nhat}_0 \, dV +
  \int_{\Gamma_2} \mu_0 \nhat_0 \cdot \ddot{\nhat}_0 \, dS \\
  &= - \int_\Omega \lambda_0 | \dot{\nhat}_0 |^2 \, dV -
  \int_{\Gamma_2} \mu_0 | \dot{\nhat}_0 |^2 \, dS ,
\end{align*}
which leads to
\begin{equation}\label{eqn:second-order}
  \delta^2 \! \calF[\nhat_0](\bmu) -
  \int_\Omega \lambda_0 | \bmu |^2 \, dV -
  \int_{\Gamma_2} \mu_0 | \bmu |^2 \, dS \ge 0 , ~~
  \forall \bmu \in \calU_0 .
\end{equation}
The above, then, is our necessary condition for local stability of
$\nhat_0$, the analogue of \cref{eqn:discrete-second-order}.  Positive
definiteness of the quadratic form in \cref{eqn:second-order}, in the
sense
\begin{equation*}
  \delta^2 \! \calF[\nhat_0](\bmu) -
  \int_\Omega \lambda_0 | \bmu |^2 \, dV -
  \int_{\Gamma_2} \mu_0 | \bmu |^2 \, dS \ge
  c \int_\Omega |\bmu|^2 \, dV , ~~
  \forall \bmu \in \calU_0 , \text{ some } c > 0 ,
\end{equation*}
would be sufficient for local stability.  Viewed in terms of
expansions, we have
\begin{equation*}
  \calF[\nhat_\eps] = \calF[\nhat_0] + \frac12 \eps^2 \biggl[
  \delta^2 \! \calF[\nhat_0](\dot{\nhat}_0) -
  \int_\Omega \lambda_0 |\dot{\nhat}_0|^2 \, dV -
  \int_{\Gamma_2} \mu_0 |\dot{\nhat}_0|^2 \, dS \biggr] + o(\eps^2) ,
\end{equation*}
for $\nhat_0\in\calN_0$ satisfying \cref{eqn:first-order}.  The
approach taken here is classical.  It is, in essence, that of
\cite[\S\S{}IV.7.2,\,IV.8.1]{courant:hilbert:53}, used in the setting
of liquid crystals in \cite[\S3.5]{virga:94}.  Similar results,
derived instead in terms of expansions, are found in
\cite{rosso:virga:kralj:04}.

\subsection{Examples}

We illustrate the application of the stability criterion above to some
examples, some of which will be considered again later in the context
of electric fields.  We first observe that when the ground state
$\nhat_0$ is uniform (which is the case in all the classical \Freed\
transitions), then the second variation of the magnetic-field model
free energy \cref{eqn:FH} takes the simple form
\begin{equation*}
  \delta^2 \! \calFH [\nhat_0](\bmu) = \int_\Omega \bigl[
  K_1 (\div\bmu)^2 + K_2 (\nhat_0\cdot\curl\bmu)^2 +
  K_3 |\nhat_0\times\curl\bmu|^2 - \chia (\bmH\cdot\bmu)^2 \bigr] \, dV .
\end{equation*}
Here we have dropped the $K_{24}$ term and the term associated with
the anchoring energy on $\Gamma_2$, since neither will appear in our
examples below.  It is also the case that if $\nhat_0$ is uniform and
in addition $\nhat_0 \perp \bmH$ (which is the case in all the
classical \Freed\ transitions with $\chia>0$), then it necessarily
follows that the associated equilibrium Lagrange multiplier field
$\lambda_0$ will be zero.

\subsubsection{Classical \Freed\ transitions}

\label{sec:classical-Freeds}

\begin{figure}
  \centering
  \subfloat[splay geometry ($\chia>0$)]{\label{fig:Freed-geoms-left}
    \includegraphics[width=.31\textwidth]{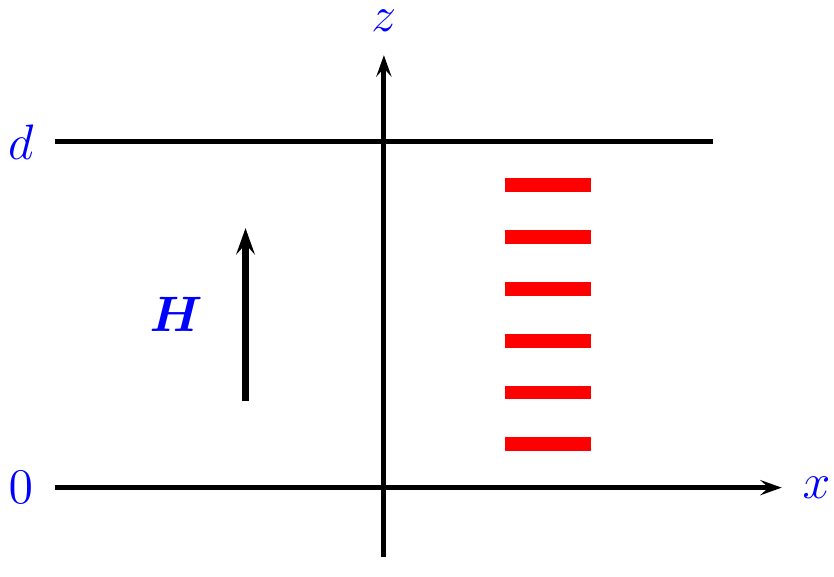}}
  \subfloat[bend geometry ($\chia>0$)]{\label{fig:Freed-geoms-center}
    \includegraphics[width=.31\textwidth]{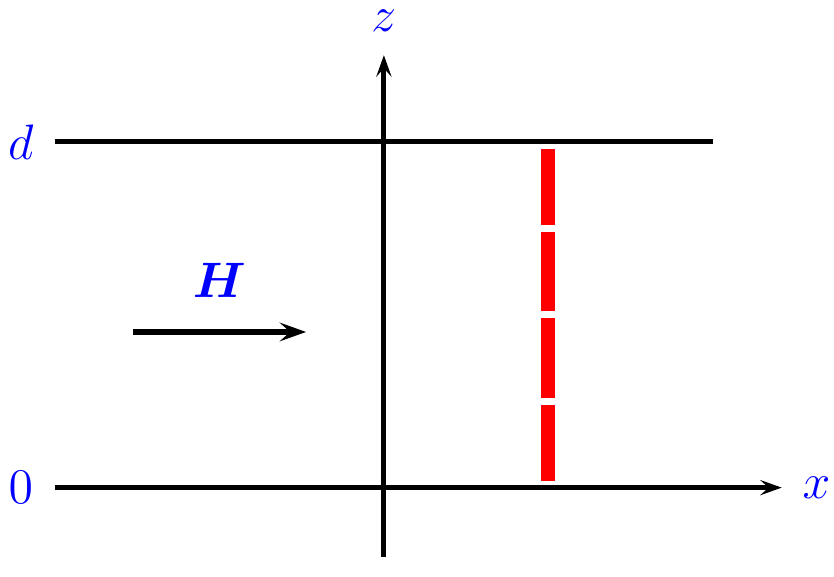}}
  \subfloat[in-plane geometry ($\chia<0$)]{\label{fig:Freed-geoms-right}
    \includegraphics[width=.31\textwidth]{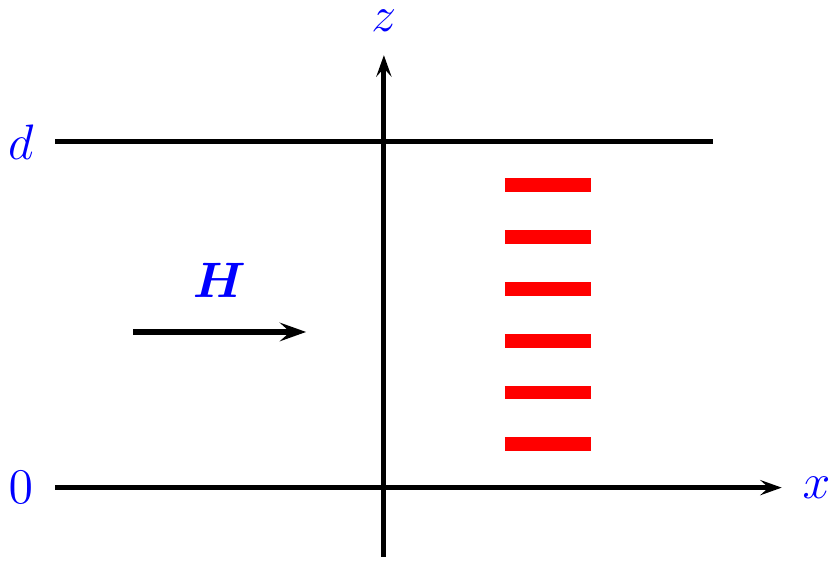}}
  \caption{Three geometries of example magnetic-field \Freed\
    transitions.  The liquid crystal film is confined to $0 < z < d$.
    The ground state equilibrium solutions $\nhat_0$ are indicated, as
    are the orientations of the magnetic fields and the signs of the
    diamagnetic anisotropy $\chia$.  Strong anchoring (Dirichlet
    boundary conditions) on the director field $\nhat$ is assumed on
    $z=0$ and $z=d$.}
  \label{fig:Freed-geoms}
\end{figure}

We consider three of the classical \Freed\ geometries, as shown in
\cref{fig:Freed-geoms}.
% \begin{figure}
%   \centering
%   \includegraphics[width=.31\textwidth]{H_splay_Freed}\hfill
%   \includegraphics[width=.31\textwidth]{H_bend_Freed}\hfill
%   \includegraphics[width=.31\textwidth]{H_neg_chia_Freed}
%   \caption{Geometries of three example magnetic-field Fr\'{e}edericksz
%     transitions.  The liquid crystal film is confined to $0 < z < d$,
%     with ground state equilibrium solutions $\nhat_0$ indicated and
%     strong anchoring (Dirichlet boundary conditions) on $\nhat$
%     assumed on $z=0$ and $z=d$.  For the left and center geometries, a
%     positive diamagnetic anisotropy is assumed ($\chia>0$); while
%     $\chia<0$ for the geometry in the figure on the right.}
%   \label{fig:Freed-geoms}
% \end{figure}
\Cref{fig:Freed-geoms-left} depicts the ``splay \Freed\ geometry.''
With the director field $\nhat$ assumed to be uniform in the lateral
directions and confined to the tilt plane spanned by $\ehat_x$ and
$\ehat_z$,
\begin{equation*}
  \nhat = n_x(z) \ehat_x + n_z(z) \ehat_z ,
\end{equation*}
the free energy (per unit cross-sectional area) is given by
\begin{equation*}
  \calF[\nhat] = \frac12 \int_0^d \bigl(
  K_1 n_{z,z}^2 + K_3 n_{x,z}^2 - \chia H^2 n_z^2 \bigr) \, dz .
\end{equation*}
Here $n_{z,z}$ denotes $\frac{d}{dz} n_z$, etc.  The Euler-Lagrange
equations are
\begin{equation*}
  K_3 n_{x,zz} + \lambda n_x = 0 , ~~
  K_1 n_{z,zz} + \bigl( \chia H^2 + \lambda \bigr) n_z = 0 , ~~
  n_x^2 + n_z^2 = 1 ,
\end{equation*}
and the ground state solution is
\begin{equation*}
  \nhat_0 = \ehat_x , ~~ \lambda_0 = 0 .
\end{equation*}
Thus, in terms of components, the ground state is $n_x=1$, $n_z=0$
(which satisfy the Euler-Lagrange equations in a trivial way), and
we note that $\nhat_0=-\ehat_x$ would work equally well.  The second
variation (with $\bmu = u(z) \ehat_x + w(z) \ehat_z$ restricted to the
same tilt plane as $\nhat$) is given by
\begin{equation*}
  \delta^2 \! \calF[\nhat_0](\bmu) = \int_0^d \bigl(
  K_1 w_z^2 + K_3 u_z^2 - \chia H^2 w^2 \bigr) \, dz .
\end{equation*}
% \begin{equation*}
%   \delta^2 \! \calF[\nhat_0](\bmu) = \int_0^d \bigl(
%   K_1 w_{,z}^2 + K_3 u_{,z}^2 - \chia H^2 w^2 \bigr) \, dz .
% \end{equation*}
For unsubscripted scalar fields, we denote $w_z = dw/dz$ (or $w_z
= \partial w / \partial z$, as the situation may require), etc.
Admissible variations ($\bmu\in\calU_0$) must satisfy $\nhat_0 \cdot
\bmu = 0$, which implies that $u=0$ and $\bmu = w(z) \ehat_z$.  The
stability condition \cref{eqn:second-order} thus becomes
% \begin{equation*}
%   \int_0^d \bigl( K_1 w_z^2 - \chia H^2 w^2 \bigr) \, dz \ge 0 ,
% \end{equation*}
\begin{equation*}
  \int_0^d \bigl( K_1 w_z^2 - \chia H^2 w^2 \bigr) \, dz \ge 0
  ~ \Leftrightarrow ~
  \frac{\chia H^2}{K_1} \le
  \frac{\int_0^d w_z^2 \, dz}{\int_0^d w^2 \, dz} ,
\end{equation*}
for all smooth $w$ such that $w(0) = w(d) = 0$.
% for all smooth $w$ such that $w(0) = w(d) = 0$, or
% \begin{equation*}
%   \frac{\chia H^2}{K_1} \le
%   \frac{\int_0^d w_z^2 \, dz}{\int_0^d w^2 \, dz} .
% \end{equation*}
The minimum of the Rayleigh quotient on the right-hand side above
(over smooth $w$ satisfying $w(0)=w(d)=0$) is $\pi^2 / d^2$, which
finally leads to
\begin{equation*}
  H \le \frac{\pi}{d} \sqrt{ \frac{K_1}{\chia} } =: \Hc ,
\end{equation*}
the correct instability threshold for this problem
\cite[(3.64)]{degennes:prost:93}, \cite[(3.126)]{stewart:04},
\cite[(4.43)]{virga:94}.

In a very similar way, the magnetic-field bend-\Freed\ transition
(depicted in \cref{fig:Freed-geoms-center}) has a ground state
\begin{equation*}
  \nhat_0 = \ehat_z , ~~ \lambda_0 = 0
\end{equation*}
and a second variation given by
\begin{equation*}
  \delta^2 \! \calF[\nhat_0](\bmu) = \int_0^d \bigl(
  K_1 w_z^2 + K_3 u_z^2 - \chia H^2 u^2 \bigr) \, dz .
\end{equation*}
With $\nhat_0 \cdot \bmu = 0$ implying $w=0$, \cref{eqn:second-order}
leads to
\begin{equation*}
  \int_0^d \bigl( K_3 u_z^2 - \chia H^2 u^2 \bigr) \, dz \ge 0 ,
%  ~~ \text{ for all } u \text{ such that } u(0) = u(d) = 0 ,
\end{equation*}
for all smooth $u$ such that $u(0)=u(d)=0$, giving
\begin{equation*}
  H \le \frac{\pi}{d} \sqrt{\frac{K_3}{\chia}} =: \Hc .
\end{equation*}
This again is the correct instability threshold
\cite[(3.64)]{degennes:prost:93}, \cite[(3.143)]{stewart:04},
\cite[\S4.2.4]{virga:94}.  Here we have again assumed that $\nhat$ is
restricted to $\operatorname{span}\{\ehat_x,\ehat_z\}$ and is uniform
in the lateral directions.  These two examples will be expanded upon
below, where we relax some assumptions; they also will be revisited
later with the systems subjected to electric fields (instead of
magnetic fields), in which case the splay transition will behave as
one would naively expect, but the bend transition will not.

A final classical \Freed\ transition, depicted in
\cref{fig:Freed-geoms-right}, illustrates the role of a non-vanishing
Lagrange multiplier field $\lambda_0$.  We again assume that $\nhat$
is restricted to $\operatorname{span}\{\ehat_x,\ehat_z\}$ and is
uniform in lateral directions, but here we now assume that $\chia<0$
(which encourages $\nhat$ to orient perpendicular to $\bmH$).  We note
that another simple distortion is possible here involving a twisting
of the director parallel to the $x$-$y$ plane, but we do not consider
this at the present time.  With our assumptions, the free energy and
Euler-Lagrange equations are given by
\begin{gather*}
  \calF[\nhat] = \frac12 \int_0^d \bigl(
  K_1 n_{z,z}^2 + K_3 n_{x,z}^2 - \chia H^2 n_x^2 \bigr) \, dz \\
  K_3 n_{x,zz} + ( \chia H^2 + \lambda ) n_x = 0 , ~~
  K_1 n_{z,zz} + \lambda n_z = 0 , ~~ n_x^2 + n_z^2 = 1 ,
\end{gather*}
with ground state
\begin{equation*}
  \nhat_0 = \ehat_x , ~~ \lambda_0 = - \chia H^2 .
\end{equation*}
The Lagrange multiplier field $\lambda_0$ is constant here, due to the
simplicity of the configuration; it need not be so in general.  The
constraint $\nhat_0\cdot\bmu=0$ gives
\begin{equation*}
  \bmu = w(z) \ehat_z ~ \Rightarrow ~
  \div\bmu = w_z , ~ \curl\bmu = \bfzero, ~ \bmH\cdot\bmu = 0 ,
\end{equation*}
so that the stability condition \cref{eqn:second-order} becomes
\begin{equation*}
  \delta^2 \! \calF[\nhat_0](\bmu) - \int_0^d \lambda_0 |\bmu|^2 \, dz =
  K_1 \! \int_0^d \! w_z^2 \,\, dz +
  \chia H^2 \! \int_0^d \! w^2 \, dz \ge 0 ,
\end{equation*}
giving
\begin{equation*}
  - \frac{\chia H^2}{K_1} \le
  \frac{\int_0^d w_z^2 \, dz}{\int_0^d w^2 \, dz} ~ \Rightarrow ~
  H \le \frac{\pi}{d} \sqrt{\frac{K_1}{-\chia\,}} =: \Hc.
\end{equation*}

\subsubsection{Periodic instabilities}

\label{sec:periodic-instabilities}

It is possible for simple systems, such as those depicted in
\cref{fig:Freed-geoms}, to exhibit instabilities with more structure,
such as periodic modulations in the plane of the liquid crystal film.
We consider two such examples: the ``stripe phase'' of Allender,
Hornreich, and Johnson \cite{allender:hornreich:johnson:87} and the
periodic instability of Lonberg and Meyer \cite{lonberg:meyer:85}.  In
both cases, we must relax the constraints we imposed in the examples
above (i.e., uniformity of the director in lateral directions and
confinement of it to a fixed tilt plane).

The stripe phase occurs in the bend-\Freed\ geometry
(\cref{fig:Freed-geoms-center}).  The ground state is as before:
\begin{equation*}
  \nhat_0 = \ehat_z , ~~ \lambda_0 = 0 .
\end{equation*}
The admissible variations ($\nhat_0\cdot\bmu=0$), however, are now
taken in the form
\begin{equation*}
  \bmu = u(y,z) \ehat_x + v(y,z) \ehat_y .
\end{equation*}
The domain $\Omega$ is taken as one periodic cell
\begin{equation*}
  \Omega = \{ (y,z) | - L < y < L , ~ 0 < z < d \} ,
\end{equation*}
with $u$ and $v$ periodic in $y$ (of period $2L$), vanishing on $z=0$
and $z=d$.  The actual periodicity of a periodic equilibrium solution
is chosen spontaneously by the system; thus $L$ would not be known
a-priori---see \cref{app:lonberg-meyer} for how this issue can be
addressed in the context of the next example.

Using the assumptions above, we obtain
\begin{equation*}
  \delta^2 \! \calF[\nhat_0](\bmu) = \int_\Omega \bigl[
  K_1 v_y^2 + K_2 u_y^2 + K_3 \bigl( u_z^2 + v_z^2 \bigr) -
  \chia H^2 u^2 \bigr] \, dA ,
\end{equation*}
leading to the stability condition
\begin{equation}\label{eqn:stripe-stability}
  \int_0^d \int_{-L}^L \bigl(
  K_2 u_y^2 + K_3 u_z^2 - \chia H^2 u^2 \bigr) \, dy \, dz +
  \int_0^d \int_{-L}^L \bigl(
  K_1 v_y^2 + K_3 v_z^2 \bigr) \, dy \, dz \ge 0 ,
\end{equation}
for all $u,v \in C^2(\Omegabar)$, periodic in $y$ (of period $2L$),
vanishing on $z=0$ and $z=d$.  It is clear by inspection that the
derivatives in $y$ and the $v$ component in general can only elevate
the value of the quadratic form, leading to the conclusion that there
can be no instability of $\nhat_0$ to a periodic-in-$y$ mode and that
the first instability encountered is the classical \Freed\ transition
\begin{equation*}
  u = \sin \frac{\pi z}{d} , ~~ v = 0 , ~~
  \Hc = \frac{\pi}{d} \sqrt{\frac{K_3}{\chia}} .
\end{equation*}
In fact, the stripe phase enters as a secondary bifurcation off the
branch of these classical solutions
\cite{allender:hornreich:johnson:87} (further explored in
\cite{golovaty:gross:hariharan:gartland:01,ryabtseva:09,sherman:04}).
In more quantitative terms, the quadratic form
\cref{eqn:stripe-stability} is diagonalized by the modes
\begin{equation*}
  f_{mn} = \exp\Bigl(i\frac{m\pi y}{L}\Bigr) \sin\frac{n\pi z}{d} , ~~
  m = 0 , \pm 1 , \pm 2 , \ldots , ~ n = 1, 2, \ldots ,
\end{equation*}
with
\begin{equation*}
  u = f_{mn} , ~~ v = 0 , ~~ \lambda_{mn} =
  K_2 \frac{m^2\pi^2}{L^2} + K_3 \frac{n^2\pi^2}{d^2} - \chia H^2
\end{equation*}
and
\begin{equation*}
  u = 0 , ~~ v = f_{mn} , ~~ \lambda_{mn} =
  K_1 \frac{m^2\pi^2}{L^2} + K_3 \frac{n^2\pi^2}{d^2} ,
\end{equation*}
with the leading instability mode corresponding to $u = f_{01}$,
$v=0$.  The reason things are so simple here is that $u$ and $v$ are
uncoupled.

The periodic instability of Lonberg and Meyer \cite{lonberg:meyer:85}
is more complicated and exhibits different behavior.  The geometry is
the splay-\Freed\ geometry (\cref{fig:Freed-geoms-left}).  With ground
state
\begin{equation*}
  \nhat_0 = \ehat_x , ~~ \lambda_0 = 0 ,
\end{equation*}
admissible variations now taken in the form
\begin{equation*}
  \bmu = v(y,z) \ehat_y + w(y,z) \ehat_z ,
\end{equation*}
and domain $\Omega$ taken to be one periodic cell (as in the previous
example), the stability condition \cref{eqn:second-order} becomes
\begin{equation}\label{eqn:lonberg-meyer-stability}
  \int_0^d \int_{-L}^L \bigl[
  K_1 ( v_y + w_z )^2 + K_2 ( v_z - w_y )^2 - \chia H^2 w^2 \bigr] \,
  dy \, dz \ge 0 ,
\end{equation}
for all $v,w \in C^2(\Omegabar)$, periodic in $y$ (with period $2L$),
vanishing on $z=0$ and $z=d$.  The fields $v$ and $w$ are coupled now,
and so the quadratic form is not diagonalized by simple Fourier
expansions.  In addition to experiments and theory presented in
\cite{lonberg:meyer:85}, one finds results in the brief note
\cite{oldano:86}; while in \cite[\S4.3]{virga:94}, the system is
studied as an example of a ``periodic Freedericks transition.''  We
present a somewhat different analysis in \cref{app:lonberg-meyer} and
summarize the main results now.

The experiments reported in \cite{lonberg:meyer:85} used polymer
liquid crystal materials, which are characterized by very elongated
``rod like'' molecular architecture and by having ``twist'' elastic
constants, $K_2$ in \cref{eqn:We}, that are small compared to their
``splay'' elastic constants, $K_1$.  For such materials, the authors
reported that the classical \Freed\ transition was preceded (at a
lower magnetic-field strength) by an instability to a solution that
was periodic in $y$, with a period chosen by the system.  Analysis of
the model formulated above confirms this.  There is a value
$\Kbar_2^* \doteq 0.303$ (which can be determined analytically) such
that for $K_2 / K_1 < \Kbar_2^*$, the uniform ground state $\nhat_0$
will become unstable to a periodic-in-$y$ solution of some period for
some $\Hp < \Hc$.  As $K_2 / K_1 \rightarrow \Kbar_2^*$,
$\Hp \rightarrow \Hc$, and the period of the instability mode becomes
infinite.  See \cref{app:lonberg-meyer} for details.

The examples in this section do not provide new information about
these systems.  They merely demonstrate consistency with known
results, using the framework that has been developed here.  Also, the
essential role in our framework of the Lagrange multiplier field, when
it is nonzero, has been made clear.  The extension of these ideas to
systems involving electric fields is taken up next.

\section{Stability criteria for electric fields}

\label{sec:stability-for-E-fields}

In extending the results of the previous section to the case of a
liquid-crystal system subjected to an electric field, we take into
account the inhomogeneous nature of the electric field (in general)
and its coupling to the director field, and we work with a model free
energy of the form \cref{eqn:FE}:
\begin{equation*}
  \calF[\nhat,\varphi] = \int_\Omega \Bigl[ \We(\nhat,\nabla\nhat) -
  \frac12 \epstensor(\nhat) \nabla\varphi \cdot \nabla\varphi \Bigr] dV +
  \int_{\Gamma_2} \Ws(\nhat) \, dS .
\end{equation*}
This now is a function of two state variables: $\nhat$ (the director
field) and $\varphi$ (the electric potential).  The dielectric tensor
${\mathlarger\epstensor}$ is as given in \cref{eqn:WE}, with
$\eperp, \epara > 0$.  It follows that for any unit-length vector
field $\nhat$, ${\mathlarger\epstensor}(\nhat)$ is real symmetric
positive definite and satisfies
\begin{equation*}
  \epsz \min \{ \eperp, \epara \}
  \int_\Omega |\nabla\varphi|^2 \, dV \le
  \int_\Omega \epstensor(\nhat)
  \nabla\varphi \cdot \nabla\varphi \, dV \le
  \epsz \max \{ \eperp, \epara \}
  \int_\Omega |\nabla\varphi|^2 \, dV .
\end{equation*}
Thus the equilibrium problem has an intrinsic minimax nature to it (as
previously observed), with stationary points of $\calF$ (subject to
$\nhat\in\calN$, $\varphi\in{\mathlarger\PHI}$) maximizing with
respect to $\varphi$, locally minimizing with respect to $\nhat$.  A
stability analysis can be developed from this point of view.  However,
we have found it more direct to employ deflation, and that is the
approach we use in what follows.

\subsection{Stability criteria}

\label{sec:deflation}

It is natural to think of the electric field as ``slaved'' to the
director field.  In the setting of liquid crystal hydrodynamics, for
example, the time scale for director orientation changes is several
orders of magnitude slower than that for changes in the electric
displacement \cite{shiyanovskii:lavrentovich:10}, enabling one to
model (at this level) the electric field as adjusting instantaneously
to changes in the director field.  Motivated by this, we define an
operator $T : \calN \rightarrow {\mathlarger\PHI}$ that
gives the unique electric potential $\varphi$ associated with a given
director field $\nhat$ via
% \begin{equation*}
%   \nhat \in \calN ~ \Rightarrow ~ \varphi = T(\nhat) ,
%   \text{ such that } \delta_\varphi \calF[\nhat,\varphi] = 0 , ~
%   \varphi \in \PHI .
% \end{equation*}
\begin{equation*}
  \nhat \in \calN ~ \Rightarrow ~ T(\nhat) = \varphi \in \PHI ,
  \text{ such that } \delta_\varphi \calF[\nhat,\varphi] = 0 .  
\end{equation*}
The weak and strong forms characterizing $\varphi$ are
\begin{equation*}
  \delta_\varphi \calF[\nhat,\varphi](\psi) = 0 , ~ \forall \psi \in \PSI_0
\end{equation*}
or
\begin{equation*}
  \int_\Omega \epstensor(\nhat) \nabla\varphi \cdot \nabla\psi \, dV
  = 0 , ~ \forall \psi \in \PSI_0  
\end{equation*}
and
\begin{equation*}
  \div \bigl[ \epstensor(\nhat) \nabla\varphi \bigr] = 0
  \text{ in } \Omega , ~~ \varphi = 0 \text{ on } \Gamma_1 , ~
  \varphi = V \text{ on } \Gamma_2 , ~
  \varphi \text{ periodic on } \Gamma_3 .
\end{equation*}
Here ${\mathlarger\PSI}_0$ is the class of admissible variations of
$\varphi$:
\begin{equation*}
  \PSI_0 = \bigl\{ \psi \in C^2(\Omegabar) \, \bigl| \,
  \psi = 0 \text{ on } \Gamma_1 \text{ and } \Gamma_2 ,
  \psi \text{ periodic on } \Gamma_3 \bigr\} .
\end{equation*}
The strong form Euler-Lagrange equation above is simply the Gauss Law
in a medium with no free charge: $\div\bmD=0$.

We define our deflated free energy using the map $T$:
\begin{equation*}
  \Ftilde[\nhat] := \calF[\nhat,T(\nhat)] .
\end{equation*}
This device is similar to that used in
\cite[\S4]{hardt:kinderlehrer:lin:86}, for example.  Our previously
established results apply without change to $\Ftilde$, giving
first-order and second-order necessary conditions for local stability
of $\nhat_0$
\begin{gather}
  \delta \Ftilde[\nhat_0](\bmv) =
  \int_\Omega \lambda_0 \nhat_0 \cdot \bmv \, dV +
  \int_{\Gamma_2} \mu_0 \nhat_0 \cdot \bmv \, dS , ~~
  \forall \bmv \in \calV_0 \label{eqn:first-order-Ftilde} \\
  \delta^2 \! \Ftilde[\nhat_0](\bmu) -
  \int_\Omega \lambda_0 | \bmu |^2 \, dV -
  \int_{\Gamma_2} \mu_0 | \bmu |^2 \, dS \ge 0 , ~~
  \forall \bmu \in \calU_0 . \label{eqn:second-order-Ftilde}
\end{gather}
To express these in terms of the original $\calF$ requires some
chain-rule calculus, for which we require the derivative $DT$ of the
map $T$.  For a given director field $\nhat_0\in\calN$ with associated
electric potential field $\varphi_0 = T(\nhat_0)$, $DT(\nhat_0)$ is
the linear transformation on $\calV_0$ to ${\mathlarger\PSI}_0$ that
gives the first-order change in $\varphi_0$ associated with a small
perturbation of $\nhat_0$.  It is most readily obtained by
substituting $\nhat = \nhat_0 + \eps \bmv$ and
$\varphi = \varphi_0 + \eps \psi$ in
$\div\bigl[{\mathlarger\epstensor}(\nhat)\nabla\varphi\bigr]=0$, which
gives the strong-form characterization of $\psi = DT(\nhat_0) \bmv$:
\begin{subequations}\label{eqn:psi-strong}
  \begin{equation}
    \div \bigl[ \epstensor(\nhat_0) \nabla\psi - \bmd_0 \bigr] = 0
    \text{ in } \Omega , ~~
    \psi = 0 \text{ on } \Gamma_1 \text{ and } \Gamma_2 , ~
    \psi \text{ periodic on } \Gamma_3 ,
  \end{equation}
  where
  \begin{equation}\label{eqn:dzero}
    \bmd_0 := \epsz \epsa (
    \nhat_0\otimes\bmv + \bmv\otimes\nhat_0 ) \bmE_0 , ~~
    \bmE_0 = - \nabla\varphi_0 .
  \end{equation}
\end{subequations}
The associated weak form is
\begin{equation*}
  \int_\Omega \bigl[ \epstensor(\nhat_0) \nabla\psi - \bmd_0 \bigr]
  \cdot \nabla\chi \, dV = 0 , ~ \forall \chi \in \PSI_0 .
\end{equation*}
We note that
\begin{equation*}
  \psi = 0 \text{ on } \Omega ~ \Leftrightarrow ~
  \div \bmd_0 = 0 \text{ on } \Omega
\end{equation*}
and
\begin{equation}\label{eqn:ddotgpsi}
  \int_\Omega \bmd_0 \cdot \nabla\psi \, dV = 
  \int_\Omega \epstensor(\nhat_0) \nabla\psi \cdot \nabla\psi \, dV ,
\end{equation}
since $\psi$ is in ${\mathlarger\PSI}_0$ as well.  It is also the case
that a $\psi$ field that is not identically zero cannot be a nonzero
constant field, by virtue of the homogeneous boundary conditions that
it must satisfy.  Thus if $\psi$ is not identically zero, then
$\nabla\psi$ cannot be identically zero either.  The term $\bmd_0$ and
the observations above play an important role in our development.

The field $\bmd_0$ admits various interpretations.  It has the
dimensions of polarization (charge per unit area) and can most
immediately be seen as the first-order change in the electric
displacement associated with the perturbation
$\nhat_0 \mapsto \nhat_0 + \eps \bmv$ (while holding the electric
field fixed):
\begin{equation*}
  \epstensor(\nhat_0+\eps\bmv) \bmE_0 =
  \bmD_0 + \eps \bmd_0 + o(\eps) , \text{ with }
  \bmD_0 = \epstensor(\nhat_0) \bmE_0 .
\end{equation*}
One can view this instead in terms of the induced polarization.  The
linear dielectric properties that underlie the basic relationship that
we have used ($\bmD = {\mathlarger\epstensor}(\nhat) \bmE$) are
\begin{equation*}
  \bmD = \epsz \bmE + \bmP , ~~
  \bmP = \epsz \chitensor(\nhat) \bmE , ~~
  \chitensor = \chiperp \bfI + ( \chipara - \chiperp ) (\nhat\otimes\nhat) .
\end{equation*}
Here $\bmP$ is the polarization (dipole moment per unit volume)
induced by the electric field, and $\chitensor$ is the relative
electric susceptibility tensor.  By definition, a linear dielectric is
one in which the polarization is a linear transform of the local
electric field, here represented by a tensor field (since the medium
is anisotropic and inhomogeneous, in general).  The relationship
between the permittivities $\eperp$ and $\epara$ and the
susceptibilities $\chiperp$ and $\chipara$ is simply
\begin{equation*}
  \eperp = 1 + \chiperp, \quad \epara = 1 + \chipara ,
\end{equation*}
which implies that $\epsa = \epara - \eperp = \chipara - \chiperp$.
Thus
\begin{equation*}
  \bmd_0 = \epsz \epsa
  ( \nhat_0\otimes\bmv + \bmv\otimes\nhat_0 ) \bmE_0 =
  \epsz ( \chipara - \chiperp )
  ( \nhat_0\otimes\bmv + \bmv\otimes\nhat_0 ) \bmE_0 ,
\end{equation*}
which can be seen as the first-order change in the induced
polarization due to the perturbation of the director field $\nhat_0
\mapsto \nhat_0 + \eps \bmv$ (again holding the electric field
constant).  The divergence of polarization acts as an effective charge
distribution in general,
\begin{equation*}
  \div \bmD = 0 , ~ \bmD = \epsz \bmE + \bmP ~ \Rightarrow ~
  \div \bmE = - \frac1{\epsz} \div \bmP ,
\end{equation*}
or in the case at hand,
\begin{equation*}
  \div \bigl[ \epstensor(\nhat_0) \nabla \psi - \bmd_0 \bigr] = 0
  ~ \Rightarrow ~
  \div \bigl[ \epstensor(\nhat_0) \nabla \psi \bigr] = \div \bmd_0 .
\end{equation*}
So $\div\bmd_0$ is the source term (load) in an anisotropic Poisson
equation with homogeneous boundary conditions.  Thus if $\div\bmd_0=0$
on $\Omega$, then $\psi=0$ on $\Omega$, and this change in induced
polarization does not cause a change in the electric potential
\emph{at first order}; whereas if $\div\bmd_0\not=0$, then
$\psi\not=0$, and the change in polarization does cause a first-order
change in the potential and in the electric field as well, since
$\nabla\psi$ can't be identically zero.  We note that $\psi$ is slaved
to $\bmv$ in much the same way that $\varphi$ is slaved to $\nhat$.

To express our equilibrium conditions in terms of $\calF$ (instead of
$\Ftilde$), we proceed as follows:
\begin{equation*}
  \Ftilde[\nhat] = \calF[\nhat,T(\nhat)] ~ \Rightarrow ~
  \delta \Ftilde[\nhat](\bmv) =
  \delta_{\nhat} \calF[\nhat,T(\nhat)](\bmv) +
  \delta_{\varphi} \calF[\nhat,T(\nhat)](DT(\nhat)\bmv) .
%  \delta_{\nhat} \calF[\nhat_0,T(\nhat_0)](\bmv) ,
\end{equation*}
By the definition of $T$, however,
$\delta_{\varphi}\calF[\nhat,T(\nhat)]=0$; so
\begin{equation*}
  \delta \Ftilde[\nhat](\bmv) =
  \delta_{\nhat} \calF[\nhat,T(\nhat)](\bmv) .
\end{equation*}
Thus the equilibrium equations, in weak and strong form, are given by
\begin{gather*}
  \delta_{\nhat} \calF[\nhat,\varphi](\bmv) =
  \int_\Omega \lambda \nhat \cdot \bmv \, dV +
  \int_{\Gamma_2} \mu \nhat \cdot \bmv \, dS , ~~
  \forall \bmv \in \calV_0 \\
  \delta_{\varphi} \calF[\nhat,\varphi](\psi) = 0 , ~~
  \forall \psi \in {\PSI}_0
\end{gather*}
and
% \begin{gather*}
%   - \div \Bigl( \dWdgn \Bigr) + \dWdn = \lambda_0 \nhat_0
%   \text{ in } \Omega , ~
%   \nhat_0 = \nhatb \text{ on } \Gamma_1 , \\
%   \Bigl( \dWdgn \Bigr) \nuhat + \dWsdn = \mu_0 \nhat_0
%   \text{ on } \Gamma_2 , ~
%   \nhat_0 \text{ periodic on } \Gamma_3 \\
%   \div \Bigl( \frac{\partial W}{\partial\nabla\varphi} \Bigr) = 0
%   \text{ in } \Omega , ~ \varphi = 0 \text{ on } \Gamma_1 , ~
%   \varphi = V \text{ on } \Gamma_2 , ~
%   \varphi \text{ periodic on } \Gamma_3 .
% \end{gather*}
\begin{equation*}
  - \div \Bigl( \dWdgn \Bigr) + \dWdn = \lambda \nhat , ~~
  \div \Bigl( \frac{\partial W}{\partial\nabla\varphi} \Bigr) = 0 , ~~
  \text{in } \Omega ,
\end{equation*}
with boundary conditions
\begin{gather*}
  \nhat = \nhatb \text{ on } \Gamma_1 , ~
  \Bigl( \dWdgn \Bigr) \nuhat + \dWsdn = \mu \nhat
  \text{ on } \Gamma_2 , ~
  \nhat \text{ periodic on } \Gamma_3 \\
  \varphi = 0 \text{ on } \Gamma_1 , ~
  \varphi = V \text{ on } \Gamma_2 , ~
  \varphi \text{ periodic on } \Gamma_3 .
\end{gather*}
The coupling between $\nhat$ and $\varphi$ is more explicit when the
partial differential equations above are written
\begin{equation*}
  - \div \Bigl( \dWedgn \Bigr) + \dWedn = \lambda \nhat +
  \epsz \epsa \bigl( \nabla\varphi \cdot \nhat \bigr) \nabla \varphi , ~~
  \div \bigl[ \epstensor(\nhat) \nabla \varphi \bigr] = 0 ,
\end{equation*}
where $\We$ is the distortional elasticity as in \cref{eqn:We} (which
depends only on $\nhat$ and $\nabla\nhat$).

The corresponding second-order conditions can be obtained as follows.
\begin{multline*}
  \delta \Ftilde[\nhat](\bmv) =
  \delta_{\nhat} \calF[\nhat,T(\nhat)](\bmv) ~ \Rightarrow ~ \\
  \delta^2 \! \Ftilde[\nhat](\bmv) = 
  \delta_{\nhat\nhat}^2 \calF[\nhat,T(\nhat)](\bmv) +
  \delta_{\nhat\varphi}^2 \calF[\nhat,T(\nhat)](\bmv,DT(\nhat)\bmv) .
\end{multline*}
The last term above admits a simple form: with $\varphi_0=T(\nhat_0)$
and $\psi=DT(\nhat_0)\bmv$,
\begin{equation*}
%  \delta_{\nhat\varphi}^2 \calF[\nhat_0,T(\nhat_0)](\bmv,DT(\nhat_0)\bmv) =
  \delta_{\nhat\varphi}^2 \calF[\nhat_0,\varphi_0)](\bmv,\psi) =
  \int_\Omega \bmd_0 \cdot \nabla\psi \, dV =
  \int_\Omega \epstensor(\nhat_0) \nabla\psi \cdot \nabla\psi \, dV ,
\end{equation*}
where $\bmd_0$ is as defined in \cref{eqn:dzero} and we have also used
the relation \cref{eqn:ddotgpsi}.  Thus
\begin{equation*}
  \delta^2 \! \Ftilde[\nhat_0](\bmv) =
  \delta_{\nhat\nhat}^2 \calF[\nhat_0,\varphi_0](\bmv) +
  \int_\Omega \epstensor(\nhat_0) \nabla\psi \cdot \nabla\psi \, dV ,
  ~~ \varphi_0 = T(\nhat_0) , ~~ \psi = DT(\nhat_0) \bmv .
\end{equation*}
We thus have the following final form of the second-order necessary
condition for local stability of the equilibrium director field
$\nhat_0$ and associated electric potential field $\varphi_0=T(\nhat_0)$:
\begin{multline}\label{eqn:second-order-electric}
  \delta_{\nhat\nhat}^2 \calF[\nhat_0,\varphi_0](\bmu) +
  \int_\Omega \epstensor(\nhat_0) \nabla\psi \cdot \nabla\psi \, dV \\
  {} - \int_\Omega \lambda_0 |\bmu|^2 \, dV -
  \int_{\Gamma_2} \mu_0 |\bmu|^2 \, dS \ge 0 , ~~
  \forall \bmu \in \calU_0 ,
\end{multline}
where $\psi=DT(\nhat_0)\bmu$ is as defined in \cref{eqn:psi-strong}.
Positive definiteness of the quadratic form above would be sufficient
for local stability of $\nhat_0$, $\varphi_0$.

\Cref{eqn:second-order-electric} differs from the magnetic-field
version \cref{eqn:second-order} only by the term involving
$\nabla\psi$, which captures the increase in the second variation of
the free energy associated with the change in the electric potential
caused by a change in the director field.  The non-negative nature of
the contribution is a direct consequence of the fact that the
equilibrium electric potential $\varphi_0 = T(\nhat_0)$ is maximizing:
\begin{equation*}
  \calF[\nhat_0,\varphi_0] = \max_{\varphi\in\PHI} \calF[\nhat_0,\varphi] .
\end{equation*}
The characterization of $\psi$ from that point of view is
\begin{gather*}
  \max_{\psi\in\Psi_0} \int_\Omega \Bigl[ \bmd_0 \cdot \nabla\psi -
  \frac12 \epstensor(\nhat_0) \nabla\psi \cdot \nabla\psi \Bigr] \, dV =
  \frac12 \int_\Omega \epstensor(\nhat_0) \nabla\psi \cdot \nabla\psi
  \, dV , \\
  \int_\Omega \bigl[ \epstensor(\nhat_0) \nabla\psi - \bmd_0 \bigr]
  \cdot \nabla\chi \, dV = 0 , ~ \forall \chi \in \PSI_0 .
\end{gather*}
The expression involving $\nabla\psi$ in
\cref{eqn:second-order-electric} can be viewed in terms of the
electric field, instead of the electric potential: $\varphi =
\varphi_0 + \eps \psi ~ \Rightarrow$
\begin{equation*}
  \nabla\varphi = \nabla\varphi_0 + \eps \nabla\psi ~ \Rightarrow ~
  \bmE = \bmE_0 + \delta\bmE , ~~
  \bmE = - \nabla\varphi , ~ \bmE_0 = - \nabla\varphi_0 , ~
  \delta\bmE = - \eps \nabla\psi .
\end{equation*}
Thus
\begin{equation*}
  \frac12 \eps^2 \int_\Omega
  \epstensor(\nhat_0) \nabla\psi \cdot \nabla\psi \, dV =
  \frac12 \int_\Omega
  \epstensor(\nhat_0) \delta\bmE \cdot \delta\bmE \, dV .
\end{equation*}
When an equilibrium director field $\nhat_0$ is perturbed
($\nhat_0 \mapsto \nhat_0 + \delta\nhat$), the associated equilibrium
electric field will be perturbed as well
($\bmE_0 \mapsto \bmE_0 + \delta\bmE$), and the expression above gives
the change in the electric contribution to the free energy associated
with this (at the level of the second variation).  The induced change
can only lead to an increase in the free energy.  An example discussed
in the next subsection gives an illustration.

Some conclusions can immediately be drawn from the local stability
criterion \cref{eqn:second-order-electric}.  Observe that if $\psi=0$
(which happens if and only if $\bmd_0$ is divergence free on
$\Omega$), then \cref{eqn:second-order-electric} is the same as
\cref{eqn:second-order} but with electric-field parameters ($\epsz$,
$\epsa$, $\bmE_0=-\nabla\varphi_0$) instead of magnetic-field
parameters ($\chia$, $\bmH$).  It follows that in such cases,
stability thresholds for electric-field \Freed\ transitions, for
example, would be given by the recipes of
\cite[\S3.3.1]{degennes:prost:93} and \cite[\S3.5]{stewart:04}, e.g.,
\begin{equation}\label{eqn:Hc-Ec-splay}
  \Hc = \frac{\pi}{d} \sqrt{\frac{K_1}{\chia}} ~ \leftrightarrow ~
  \Ec = \frac{\pi}{d} \sqrt{\frac{K_1}{\epsz\epsa}}
\end{equation}
for the electric-field splay-\Freed\ transition, as analyzed in
\cite{deuling:72} and \cite[\S3.5]{stewart:04}.  In the common
alternate notation $\chia=\mu_0\Delta\chi$ (with $\mu_0$ the
free-space magnetic permeability) and $\epsa=\Delta\eps$, the formulas
above would essentially be ``carbon copies'' of each other.  In the
examples below, we shall see that indeed $\div\bmd_0=0$ in this
case of the electric-field splay transition.
If, on the other hand, $\psi\not=0$, then the contribution of the
$\nabla\psi$ term to the left hand side of
\cref{eqn:second-order-electric} will be strictly positive and will
necessarily elevate the electric-field \Freed\ threshold compared to
the formulas given in \cite[\S3.3.1]{degennes:prost:93} and
\cite[\S3.5]{stewart:04}.  This will be seen to be the case in both
the electric-field bend-\Freed\ transition (with $\epsa>0$) and the
electric-field splay-\Freed\ transition (with $\epsa<0$).  The
``litmus test,'' then, is whether or not $\div\bmd_0=0$, i.e., whether
or not
\begin{equation*}
  \div \bigl[ ( \nhat_0\otimes\bmu + \bmu\otimes\nhat_0 ) \bmE_0 \bigr]
  = 0 \text{ on } \Omega ,
\end{equation*}
for all admissible variations $\bmu \in \calU_0$.

\subsection{Examples}

The simple test of whether $\div\bmd_0$ is zero or not can be used,
for example, to identify which of the classical \Freed\ transitions
can be expected to differ qualitatively in the electric-field case
from the magnetic-field case.  Consider first the electric-field
splay-\Freed\ transition, as depicted in \cref{fig:Freed-geoms-left}
but with an electric field instead of a magnetic field (and $\epsa>0$
instead of $\chia>0$)---the electric field is generated by electrodes
at the top and bottom of the liquid crystal cell held at a constant
potential difference by an external variable voltage source (as
pictured in \cref{fig:domain}).  In this case, the ground state is
given by
\begin{equation*}
  \nhat_0 = \ehat_x , ~~ \bmE_0 = E_0 \ehat_z ,
\end{equation*}
and the admissible variations (confined to the tilt plane spanned by
$\ehat_x$ and $\ehat_z$) are
\begin{equation*}
  \nhat_0 \cdot \bmu = 0 ~ \Rightarrow ~ \bmu = w(z) \ehat_z ,
\end{equation*}
from which we obtain
\begin{equation*}
  \bmd_0 = \epsz \epsa ( \nhat_0\otimes\bmu + \bmu\otimes\nhat_0 )
  \bmE_0 = \epsz \epsa E_0 w(z) \ehat_x ~ \Rightarrow ~
  \div \bmd_0 = 0 .
\end{equation*}
Thus the electric-field coupling will not effect the \Freed\
threshold, and the recipe of \cite[\S3.3.1]{degennes:prost:93} and
\cite[\S3.5]{stewart:04} will give the correct result
\cref{eqn:Hc-Ec-splay}.
% \begin{equation*}
%   \Hc = \frac{\pi}{d} \sqrt{\frac{K_1}{\chia}}
%   ~~ \leftrightarrow ~~
%   \Ec = \frac{\pi}{d} \sqrt{\frac{K_1}{\epsz\epsa}} .
% \end{equation*}
This is consistent with \cite{deuling:72} and \cite[\S3.5]{stewart:04}.

Consider, on the other hand, the electric-field bend-\Freed\
transition, as depicted in \cref{fig:Freed-geoms-center}, again with
an electric field instead of a magnetic field.  We note that in this
geometry, the electrodes must be placed on the left and right ends of
the cell (at a sufficient separation relative to the cell gap so as to
render boundary effects negligible).  This makes these experiments
more difficult to conduct (because of the larger voltages required)
and also complicates the modeling and analysis.  The test with
$\div\bmd_0$ is still easy to apply.  With ground state and variations
given by
\begin{equation*}
  \nhat_0 = \ehat_z , ~~
  \bmE_0 = E_0 \ehat_x , ~~
  \bmu = u(z) \ehat_x ,
\end{equation*}
we obtain
\begin{equation*}
  \bmd_0 = \epsz \epsa E_0 u(z) \ehat_z ~ \Rightarrow ~
  \div \bmd_0 = \epsz \epsa E_0 u_z .
\end{equation*}
Since $\div\bmd_0$ is not necessarily zero, we anticipate an elevated
instability threshold for the electric field compared to the formula
obtained using the magnetic-field analogy.  It is shown in
\cite{arakelyan:karayan:chilingaryan:84} (also derived below) that
this is indeed the case, with
\begin{equation}\label{eqn:Ec-bend}
  \Ec = \sqrt{\frac{\epara}{\eperp}} \times \frac{\pi}{d}
  \sqrt{\frac{K_3}{\epsz\epsa}} .
%  \frac{\pi}{d} \sqrt{\frac{K_3}{\epsz\epsa}} .
\end{equation}
The elevating factor $\sqrt{\epara/\eperp}$ above is not necessarily
small.  For example, using values from \cite[Table\,D.3]{stewart:04}
for the material 5CB near $26^{\circ}$C, we have
\begin{equation*}
  \epara = 18.5 , ~ \eperp = 7 ~\Rightarrow~
  \sqrt{\frac{\epara}{\eperp}} \doteq 1.63 ,
\end{equation*}
which implies a 63\% higher switching voltage.  Such a factor
($\epara/\eperp$) has appeared in investigations of
electric-field-induced instabilities in other systems as well---see
for example \cite{bevilacqua:napoli:05}.

Another case that manifests such behavior is the electric-field
splay-\Freed\ transition with $\epsa<0$, that is,
$0 < \epara < \eperp$.  This is as depicted in
\cref{fig:Freed-geoms-right}, but with $\bmE$ instead of $\bmH$.
With $\nhat$ still restricted to $\operatorname{span}
\{\ehat_x,\ehat_z\}$, we have
\begin{equation*}
  \nhat_0 = \ehat_x , ~
  \bmE_0 = E_0 \ehat_x , ~
  \bmu = w(z) \ehat_z ~ \Rightarrow ~
  \bmd_0 = \epsz \epsa E_0 w(z) \ehat_z ~ \Rightarrow ~
  \div\bmd_0 = \epsz \epsa E_0 w_z .
\end{equation*}
In this case, it is shown in \cite{arakelyan:karayan:chilingaryan:84}
that
\begin{equation}\label{eqn:Ec-splay}
  \Ec = \sqrt{\frac{\eperp}{\epara}} \times \frac{\pi}{d}
  \sqrt{\frac{-K_1}{\epsz\epsa}} .
\end{equation}

Of the six classical electric-field \Freed\ transitions (three with
$\epsa>0$, three with $\epsa<0$), the two identified above are the
only ones that exhibit this anomalous behavior.  While one might guess
at first that all geometries with in-plane electric fields might give
$\div\bmd_0\not=0$, that proves not to be the case.  Both of the
twist-\Freed\ transitions have $\div\bmd_0=0$:
\cref{fig:Freed-geoms-right} with $\epsa<0$ and
$\nhat\in\operatorname{span}\{\ehat_x,\ehat_y\}$ and the transition
(which is not depicted) with $\nhat_0=\ehat_y$, $\bmE_0=E_0\ehat_x$,
$\epsa>0$, $\nhat\in\operatorname{span} \{\ehat_x,\ehat_y\}$.  In the
latter case, for example, we have
\begin{equation*}
  \nhat_0 = \ehat_y , ~
  \bmE_0 = E_0 \ehat_x , ~
  \bmu = u(z) \ehat_x ~ \Rightarrow ~
  \bmd_0 = \epsz \epsa E_0 u(z) \ehat_y ~ \Rightarrow ~
  \div \bmd_0 = 0 .
\end{equation*}

A natural question is what is it, from a physical point of view, that
distinguishes these two cases.  Consider, for example, the
electric-field bend-\Freed\ transition with $\epsa>0$ (the second
example discussed above, which has the elevated threshold
\cref{eqn:Ec-bend}) versus the electric-field splay transition with
$\epsa>0$ (the first example discussed above, which has the
non-elevated threshold \cref{eqn:Hc-Ec-splay}).  In both of these
examples, there are changes in the induced polarization:
$\bmd_0\not=\bfzero$.  In the former case, however, $\div\bmd_0\not=0$
(which implies $\psi\not=0$ and $\nabla\psi\not=\bfzero$); whereas in
the latter case, $\div\bmd_0=0$ (and $\psi=0$).  Thus while both
systems experience changes in the equilibrium electric field
accompanying a perturbation in the equilibrium director field, in the
former case, this change in $\bmE$ comes at \emph{first order}
($\bmE = \bmE_0 + \delta \bmE$,
$\delta \bmE = - \eps \nabla \psi \not= \bfzero$), while in the latter
case, the change comes at a higher order
($\delta\bmE = - \eps \nabla \psi = \bfzero$).  The difference between
the two cases comes down to peculiarities of the coupling between
$\delta\nhat$ and $\delta\bmE$.

In the two cases for which we have a non-vanishing $\div\bmd_0$, in
order to derive the formulas for the elevated switching thresholds
given above in \cref{eqn:Ec-bend} and \cref{eqn:Ec-splay} using our
stability criterion \cref{eqn:second-order-electric}, it is necessary
to evaluate the term involving $\nabla\psi$.  We now show how this
can be done for the case of the electric-field bend-\Freed\ transition
(with $\epsa>0$), modulo some simplifying assumptions.

For the electric-field bend-\Freed\ transition, as depicted in
\cref{fig:Freed-geoms-center} (but with $\bmH$ replaced by $\bmE$
and $\epsa>0$), we consider the behavior in the interior of the cell,
sufficiently removed from boundary influences at the left and right
boundaries that we can accept the simplifying assumptions
\begin{equation}\label{eqn:bend-nhat-E}
  \nhat = n_x(z) \ehat_x + n_z(z) \ehat_z , \quad
  \bmE = E_x(z) \ehat_x + E_z(z) \ehat_z .
\end{equation}
We are, in essence, looking at an ``outer solution'' (in the sense of
singular perturbations and boundary layer theory).  We express the
free energy in terms of $\bmE$ (instead of $\varphi$) and employ a
more convenient representation for the electric-field contribution:
\begin{equation*}
  \calF[\nhat,\bmE] = \int_0^d W(\nhat,\nabla\nhat,\bmE) \, dz ,
\end{equation*}
with
\begin{align*}
  2W &= K_1 (\div\nhat)^2 + K_2 (\nhat\cdot\curl\nhat)^2 +
        K_3 |\nhat\times\curl\nhat|^2 - \epsz \bigl[
        \eperp |\nhat\times\bmE|^2 + \epara (\nhat\cdot\bmE)^2 \bigr] \\
     &= K_1 n_{z,z}^2 + K_3 n_{x,z}^2 - \epsz \bigl[
        \eperp (n_zE_x-n_xE_z)^2 + \epara (n_xE_x+n_zE_z)^2 \bigr] .
\end{align*}
The Euler-Lagrange equations for $\nhat$ are given by
\begin{align*}
  K_3 n_{x,zz} + \epsz [
  \eperp (n_xE_z-n_zE_x) E_z + \epara (n_xE_x+n_zE_z) E_x \bigr] +
  \lambda n_x &= 0 \\
  K_1 n_{z,zz} + \epsz [
  \eperp (n_zE_x-n_xE_z) E_x + \epara (n_xE_x+n_zE_z) E_z \bigr] +
  \lambda n_z &= 0 ,
\end{align*}
subject to $n_x^2+n_z^2=1$ and boundary conditions $n_x(0)=n_x(d)=0$,
$n_z(0)=n_z(d)=1$, with ground state solution given by
\begin{equation*}
  \nhat_0 = \ehat_z , ~~
  \bmE_0 = E_0 \ehat_x , ~~
  \lambda_0 = - \epsz \eperp E_0^2 .
\end{equation*}
The stability criterion \cref{eqn:second-order-electric} requires
$\delta_{\nhat\nhat}^2 \calF[\nhat_0,\varphi_0](\bmu)$, which can be
expressed in the following form when $\nhat_0=\text{const}$:
\begin{multline*}
  \delta_{\nhat\nhat}^2 \calF[\nhat_0,\bmE_0](\bmu) =
  \int_\Omega \Bigl\{
  K_1 (\div\bmu)^2 + K_2 (\nhat_0\cdot\curl\bmu)^2 +
  K_3 |\nhat_0\times\curl\bmu|^2 \\
  {} - \epsz \bigl[
  \eperp |\bmu\times\bmE_0|^2 + \epara (\bmu\cdot\bmE_0)^2 \bigr]
  \Bigr\} \, dV .
\end{multline*}
In the present case (with $\bmu=u(z)\ehat_x$), this becomes
\begin{equation*}
  \delta_{\nhat\nhat}^2 \calF[\nhat_0,\bmE_0](\bmu) = \int_0^d
  \bigl( K_3 u_z^2 - \epsz \epara E_0^2 u^2 \bigr) \, dz .
\end{equation*}
Observe that if $\bmd_0$ were divergence free (and $\psi$ identically
zero), then the stability condition \cref{eqn:second-order-electric}
would become
\begin{equation*}
  \delta_{\nhat\nhat}^2 \calF[\nhat_0,\bmE_0](\bmu) -
  \int_0^d \lambda_0 |\bmu|^2 \, dz = \int_0^d \bigl(
  K_3 u_z^2 - \eps_a \epsa E_0^2 u^2 \bigr) \, dz \ge 0 ,
\end{equation*}
for all smooth $u$ satisfying $u(0)=u(d)=0$.  Here we have used
$\lambda_0=-\epsz\eperp E_0^2$ and $|\bmu|^2=u^2$.  This would give
\begin{equation*}
  E_0^2 \le \frac{K_3}{\epsz\epsa}
  \frac{\int_0^d u_z^2 \, dz}{\int_0^d u^2 \, dz} ~ \Rightarrow ~
  \Ec = \frac{\pi}{d} \sqrt{\frac{K_3}{\epsz\epsa}} ,
\end{equation*}
which is the value that the magnetic-field analogy of
\cite[\S3.3.1]{degennes:prost:93} and \cite[\S3.5]{stewart:04} would
predict.

To determine the contribution to \cref{eqn:second-order-electric} from
$\nabla\psi$, it is convenient to interpret the expression in terms of
the electric field rather than the electric potential.  First note
that with the electrodes at the left and right ends of the cell, the
upper and lower boundaries of the liquid-crystal film would just be
glass substrates (typically with other dielectric layers, such as
polymer alignment layers, polarizers, and the like).  Thus the
electric field would extend above and below the liquid crystal layer
(into $z>d$ and $z<0$).  Next, with our simplified modeling
assumptions \cref{eqn:bend-nhat-E}, the basic relations from the
electrostatic Maxwell equations give
\begin{equation*}
  \curl\bmE = \bfzero , ~ \div\bmD = 0 ~ \Rightarrow ~
  E_x = \text{const} , ~ D_z = \text{const} .
\end{equation*}
The constants can be determined as follows.  Assuming that all
dielectric interfaces (liquid-crystal/polymer, polymer/glass,
glass/air, etc.)\ are planar and parallel to the $x$-$y$ plane, then
the quantities $E_x$ and $D_z$ would be continuous across these
interfaces and would continue to hold with the same constant values
above and below the cell (since tangential components of the electric
field and normal components of the electric displacement are
continuous across material interfaces in general).  Assuming also that
the electrodes are sufficiently tall that we can model them as having
infinite extent in the $\pm z$ directions, then we would have that
\begin{equation*}
  \bmE \rightarrow E_0 \ehat_x , ~ D_z \rightarrow 0 ,
  \text{ as } z \rightarrow \pm \infty .
\end{equation*}
We conclude that the $E_x$ and $D_z$ constants are $E_0$ and $0$, so
that
\begin{equation*}
  \bmE = E_0 \ehat_x + E_z(z) \ehat_z , ~~ D_z = 0 .
\end{equation*}

Next, recall that $\psi$ is the first-order change in the electric
potential ($\varphi=\varphi_0+\eps\psi+o(\eps)$) associated with a
small perturbation of the director field ($\nhat=\nhat_0+\eps\bmu$) in
the electrostatic equation $\div\bmD=\div[\epstensor(\nhat)\bmE]=0$,
$\bmE=-\nabla\varphi$.  Thus $-\eps\nabla\psi$ is the associated
first-order change in the electric field: $\bmE=\bmE_0+\delta\bmE$,
$\bmE_0=-\nabla\varphi_0$, $\delta\bmE=-\eps\nabla\psi$.  In our
setting, however, $\div\bmD=0$ collapses to $D_z=0$, which is given by
\begin{equation*}
  \eperp E_z + \epsa ( n_x E_0 + n_z E_z ) n_z = 0 .
\end{equation*}
Thus to determine $\delta\bmE$ in our model, we can simply substitute
$\nhat=\nhat_0+\eps\bmu$ ($n_x=\eps u$, $n_z=1$) above and solve for
$E_z$ to conclude
\begin{equation*}
  \nabla \psi = \frac{\epsa}{\epara} E_0 u \ehat_z ,
\end{equation*}
and we obtain
\begin{equation*}
  \epstensor(\nhat_0) \nabla\psi \cdot \nabla\psi =
  \epsz \frac{\epsa^2}{\,\epara} E_0^2 u^2 .
\end{equation*}
Substituting this expression into our stability condition
\cref{eqn:second-order-electric}, we obtain
\begin{multline*}
  \delta_{\nhat\nhat}^2 \calF[\nhat_0,\bmE_0](\bmu) +
  \int_0^d \epstensor(\nhat_0) \nabla\psi \cdot \nabla\psi \, dz -
  \int_0^d \lambda_0 |\bmu|^2 \, dz = \\
  \int_0^d \Bigl( K_3 u_z^2 -
  \frac{\eperp}{\epara} \epsz \epsa E_0^2 u^2 \Bigr) \, dz \ge 0 ,
\end{multline*}
for all smooth $u$ satisfying $u(0)=u(d)=0$, giving
\begin{equation*}
  E_0^2 \le \frac{\epara}{\eperp} \frac{K_3}{\epsz\epsa}
  \frac{\int_0^d u_z^2 \, dz}{\int_0^d u^2 \, dz} ~ \Rightarrow ~
  \Ec^2 = \frac{\epara}{\eperp} \frac{K_3}{\epsz\epsa}
  \frac{\pi^2}{d^2} ,
\end{equation*}
as in \cref{eqn:Ec-bend}.  We note that this expression for $\Ec$
agrees with \cite{arakelyan:karayan:chilingaryan:84} and with
\cite[\S5.2]{richards:06} (where it is confirmed via numerics and a
perturbation expansion of the bifurcation point).  Another anomaly
exhibited by this particular system is that the \Freed\ transition can
be first order, instead of second order, and this is established in
\cite{arakelyan:karayan:chilingaryan:84,frisken:palffy:89b,%
  frisken:palffy:89,frisken:palffy:89c} and
\cite[\S5.2]{richards:06}.

\section{Conclusions}

\label{sec:conclusions}

We have considered macroscopic models of Oseen-Frank type for the
orientational properties of a material in the simplest liquid crystal
phase, an achiral uniaxial nematic liquid crystal, subjected to either
a magnetic field or an electric field, and we have developed general
criteria for the local stability of equilibrium fields.  In the case
of a system with an electric field, the stability criterion takes into
account the coupling between the director field and the electric field
(which is in general inhomogeneous) and the mutual influence that
these fields have on each other.  We have restricted our attention to
the situation in which the electric field is produced by electrodes
held at constant potential by an external voltage source, which is by
far the most common case in experiments and devices involving such
materials.

The assessment of local stability is complicated by several factors,
including the coupling between the electric field and the director
field, the inhomogeneity of the electric field, the minimax nature of
the equilibrium problem, and the pointwise unit-length constraint on
the director field.  Our general results provide a full explanation of
formulas found in \cite{arakelyan:karayan:chilingaryan:84}, here given
in \cref{eqn:Ec-bend} and \cref{eqn:Ec-splay}, and they put ideas of
\cite{rosso:virga:kralj:04} in a different context and extend them
from the case of instabilities caused by magnetic fields to
electric-field-induced instabilities, with the full coupling between
the director field and the electric field taken into account.

Our development proceeded in two stages: first for systems with
magnetic fields, followed by the analysis of systems with electric
fields.  The stability criteria in the former case mimic results from
equality-constrained optimization theory in $\Rn$; while the latter
case was reduced to the former by the use of deflation, treating the
electric field as slaved to the director field (leading to a model
that is in essence a PDE-constrained optimization problem).

A main result is the stability criterion
\cref{eqn:second-order-electric}, which extends similar results for
magnetic fields to the fully coupled electric-field case.  There, the
one-sided, stabilizing nature of the coupling is revealed: the
presence of the electric field can only elevate (never lower) an
instability threshold, compared to the threshold one would calculate
if one ignored the mutual influence between the director field and the
electric field and instead treated the electric field as a uniform
external field (analogous to the situation with a magnetic field).
Another important result is the simple test of whether or not
$\div\bmd_0=0$ (with $\bmd_0$ as in \cref{eqn:dzero}), which tells us
whether or not the electric-field coupling will play a role in
determining instability thresholds in particular systems.

From a physical point of view, the mechanism that drives the effect of
the electric-field coupling on stability thresholds is the change in
the induced polarization associated with a small perturbation of an
equilibrium director field.  The coupling has an effect on an
instability threshold when such a perturbation of the director field
causes a first-order change in the electric field.  If a perturbation
of an equilibrium director field causes a change in the electric field
of higher order, then the coupling will not affect the threshold.
This latter scenario is the more common one, and for this reason,
scientists have believed for a long time that the instability
thresholds with electric fields should be given by the same formulas
as for magnetic fields (with electric parameters simply replacing
their magnetic counterparts), as explicitly stated in standard
references.

We have presented several examples illustrating the application of the
stability criteria in settings involving \Freed\ transitions (the
classic, textbook liquid crystal instability) and also with systems
that develop periodic instabilities.  The results are consistent in
all cases with results in the literature, and they correct mistakes
found in some standard texts.  The periodic instability of Lonberg and
Meyer \cite{lonberg:meyer:85} is interesting in its own right, and we
have presented a partial analysis of it in \cref{app:lonberg-meyer}.
While working with director fields $\nhat$, the pointwise constraint
$|\nhat|=1$, and the associated Lagrange multipliers $\lambda$ and
$\mu$ is in some sense more complicated than using representations in
terms of orientation angles, once the analysis has been sorted out (as
we have done here, in a fashion), the application of the criteria to
specific systems can be cleaner and simpler than that employing
orientation angles, as our examples have illustrated.

While we have focused on models of somewhat simple systems (achiral
uniaxial nematic liquid crystals with magnetic fields or electric
fields), the approach is broader and more general and can be extended
to other phases (chiral nematics or cholesterics, smectics, etc.)\ and
to include other effects, such as flexoelectricity, ferroelectricity,
and the like.  For example, in \cref{app:flexoelectric}, we show how
one can incorporate flexoelectric effects into the theory.  In that
same appendix, we also show that the flexoelectric terms incorporated
into the free energy have no effect on any of the classical \Freed\
thresholds, though it is known that they do affect equilibrium
configurations beyond the instability thresholds.

In a completely analogous manner, stability criteria could be
developed for mesoscopic continuum models of such materials (such as
tensor-order-parameter models of Landau-de\,Gennes type).  The
coupled-electric-field models would retain the minimax nature of the
equilibrium characterization and the one-sided nature of the
instability threshold assessment (capable only of elevation).  The
state variables and constraints for such models would of course differ
from those for the macroscopic models we have considered here.

From the point of view of numerical modeling, the stability criteria
developed here have natural, implementable discrete analogues.  For
example, in \cite{gartland:ramage:12}, the stability condition
analogous to \cref{eqn:second-order-electric} takes the form of an
inequality on the minimum eigenvalue of a matrix built from the blocks
of a discretization matrix for a liquid-crystal director model:
\begin{equation*}
  \lambda_{\min} \bigl[ Z^T \! \bigl( A + D C^{-1} \! D^T \bigr) Z \bigr] \ge 0 .
\end{equation*}
Here the matrix $A+DC^{-1}\!D^T\!$ represents a certain Schur
complement associated with a deflated Hessian matrix (analogous to the
second variation of the deflated free energy we have used in
\cref{sec:deflation}), and the rectangular matrix $Z$ represents the
projection transverse to discrete directors (the discrete analogue of
the tensor field $\bfP(\nhat)$ used in $\bmu = \bfP(\nhat) \bmv$ in
our continuum setting).

\appendix

\section{Periodic instability of Lonberg and Meyer}
\label{app:lonberg-meyer}
As discussed in \cref{sec:periodic-instabilities}, the periodic
instability studied in \cite{lonberg:meyer:85} concerns a system in
the splay-\Freed\ geometry, that is, a thin-film liquid-crystal cell
with strong parallel planar anchoring on the substrates and a magnetic
field perpendicular to the substrates, as depicted in
\cref{fig:Freed-geoms-left}.  As in that figure, we adopt a fixed
Cartesian coordinate system with the $x$ and $y$ coordinates in the
plane of the film (which is assumed to be infinite) and the $z$
coordinate across the film gap ($0 < z < d$).  Thus
$\bmH = H \ehat_z$, and we impose the boundary condition
$\nhat = \ehat_x$ on $z=0$ and $z=d$.  For sufficiently weak magnetic
fields, the stable ground state is the uniform configuration
$\nhat = \nhat_0 = \ehat_x$.  The classical \Freed\ transition occurs
at a critical magnetic-field strength at which the uniform ground
state becomes unstable to a configuration with the liquid-crystal
director orienting towards the direction of the magnetic field in the
interior of the cell: $\nhat = \nhat(z)$.  For the materials used in
the experiments in \cite{lonberg:meyer:85}, the authors reported that
this transition was preceded (at a lower magnetic-field strength) by
an instability to a solution that was periodic in $y$, with a period
chosen by the system: $\nhat = \nhat(y,z)$, $2L$ periodic in $y$ ($L$
not known a-priori).  The materials used in \cite{lonberg:meyer:85}
were polymer liquid crystals, which are distinguished by having very
elongated molecular architectures and by having twist elastic
constants ($K_2$ in \cref{eqn:We}) that are small compared to their
splay elastic constants ($K_1$).  Both the classical and the periodic
solutions are assumed to be uniform in the $x$ direction.  This system
is discussed as an example of a ``periodic Freedericks transition'' in
\cite[\S4.3]{virga:94}.  We model it as follows.

Let $\calF$ denote the free energy (per unit length in $x$) of a
single periodic cell:
\begin{equation*}
  \calF[\nhat] = \int_\Omega W( \nhat, \nabla\nhat ) \, dA , \quad
  \Omega = \{ (y,z) | -L < y < L , 0 < z < d \} ,
\end{equation*}
where the free-energy density is given by
\begin{equation*}
  2 W = K_1 (\div\nhat)^2 + K_2 (\nhat\cdot\curl\nhat)^2 +
        K_3 |\nhat\times\curl\nhat|^2 - \chia (\bmH\cdot\nhat)^2 .
\end{equation*}
Here the diamagnetic anisotropy $\chia$ is assumed to be positive (as
are the elastic constants $K_1$, $K_2$, $K_3$), and
$\bmH = H \ehat_z$, with $H=\text{const}$.  For given parameters
$K_1$, $K_2$, $K_3$, $\chia$, and $H$, the optimal period of a
periodic solution is the one that minimizes with respect to $L$ the
free energy averaged over one period:
\begin{equation*}
  \min_{\nhat,L} \calF_L[\nhat] , \quad \calF_L := \frac1L \calF .
\end{equation*}
The minimization with respect to $\nhat$ is subject to the boundary
conditions, the periodicity conditions, and the pointwise unit-length
constraint $|\nhat|=1$.  We note that periodic solutions have an
arbitrary phase, which leads to one-parameter families of minimizers.
One should add a ``phase condition'' to determine a locally isolated
representative.

The uniform ground state $\nhat = \nhat_0$ satisfies the
Euler-Lagrange equations with the Lagrange multiplier field associated
with the constraint $|\nhat|=1$ equal to zero ($\lambda_0 = 0$); so
the stability of $\nhat_0$ is indicated by
$\delta^2\!\calF[\nhat_0](\bmu)$, with
$\bmu = v(y,z) \ehat_y + w(y,z) \ehat_z$, which is given by
\cref{eqn:lonberg-meyer-stability}:
\begin{equation}\label{eqn:app_d2F}
  \delta^2\!\calF[\nhat_0](\bmu) = \! \int_\Omega \bigl[
  K_1 ( v_y + w_z )^2 + K_2 ( v_z - w_y )^2 - \chia H^2 w^2
  \bigr] \, dA .
\end{equation}
We note that this agrees with the expression given in \cite[p.\,719,
col.\,2]{lonberg:meyer:85} and \cite[(4.76)]{virga:94}.  The
stationary points of \cref{eqn:app_d2F} subject to
$\int_\Omega ( v^2 + w^2 ) \, dA = \text{const}$ satisfy
\begin{equation}
  \label{eqn:app_vwEL}
  \begin{gathered}
    K_1 (v_y+w_z)_y + K_2 (v_z-w_y)_z + \lambda v = 0 , \\
    K_2 (w_y-v_z)_y + K_1 (v_y+w_z)_z + \chia H^2 w + \lambda w = 0 ,
  \end{gathered}
\end{equation}
for which any nontrivial solution $v$, $w$, $\lambda$ (subject to
homogeneous boundary conditions and periodicity conditions on $v$ and
$w$) satisfies
\begin{equation*}
  \lambda = \frac{\displaystyle
    \int_\Omega \bigl[ K_1 ( v_y + w_z )^2 +
    K_2 ( v_z - w_y )^2 - \chia H^2 w^2 \bigr] \, dA}{\displaystyle
    \int_\Omega ( v^2 + w^2 ) \, dA} .
\end{equation*}
Thus the sign of the eigenvalue $\lambda$ indicates the stability or
instability of the mode ($\lambda>0$ corresponding to
$\delta^2\!\calF[\nhat_0](\bmu)>0$ and implying local stability of
$\nhat_0$ to such a perturbation, $\lambda<0$ indicating instability).
We note that in the special case $K_1=K_2$, the equations
\cref{eqn:app_vwEL} decouple.  The case of interest, however, is
$0 < K_2 < K_1$ (when twist distortion is cheap compared to splay
distortion).

We employ the following representations for $v$ and $w$:
\begin{equation}
  \label{eqn:app_vw-series}
  \begin{aligned}
    v(y,z) &= \ahat_0(z) + \sum_{k=1}^{\infty} \Bigl[
    \ahat_k(z)\cos\frac{k\pi y}{L} + \bhat_k(z)\sin\frac{k\pi y}{L} \Bigr] \\
    w(y,z) &= \chat_0(z) + \sum_{k=1}^{\infty} \Bigl[
    \chat_k(z)\cos\frac{k\pi y}{L} + \dhat_k(z)\sin\frac{k\pi y}{L} \Bigr] .
  \end{aligned}
\end{equation}
The uniform-in-$y$ modes in \cref{eqn:app_vwEL} are given by either of
the following:
\begin{gather*}
  v = \ahat_0 = \sin \frac{l\pi z}{d} , ~~ w = 0 , ~~
  \lambda_l = K_2 \frac{l^2\pi^2}{d^2} , ~~ \text{or} \\
  v = 0 , ~~ w = \chat_0 = \sin \frac{l\pi z}{d} , ~~
  \lambda_l = K_1 \frac{l^2\pi^2}{d^2} - \chia H^2 ,
\end{gather*}
with $l = 1, 2, \ldots$~.  The latter solution pair (with $l=1$) gives
the classical stability threshold:
\begin{equation}\label{eqn:app_Hc}
  \Hc := \frac{\pi}{d} \sqrt{\frac{K_1}{\chia}} .
\end{equation}
Before embarking on a systematic consideration of the stability
eigenvalue problem for periodic-in-$y$ modes, we illustrate what
information can be obtained from a simple approximation.

We wish to know how small $K_2$ must be compared to $K_1$ in order for
a periodic mode to become unstable for $H < \Hc$, i.e., for a periodic
instability to precede the classical magnetic-field splay-\Freed\
transition.  An approximate $v$, $w$ pair that has the appropriate
symmetry (but does not satisfy \cref{eqn:app_vwEL}) is
\begin{equation}\label{eqn:app_vw-approx}
  v = A \cos\frac{\pi y}{L} \sin\frac{2\pi z}{d} , ~~
  w = B \sin\frac{\pi y}{L} \sin\frac{\pi z}{d} , ~~
  A, B \text{ const} .
\end{equation}
Substituting these into \cref{eqn:app_d2F} leads to a quadratic form
in $A$, $B$:
\begin{equation*}
  \alpha A^2 + 2 \beta A B + \gamma B^2 ,
\end{equation*}
with
\begin{equation*}
  \alpha = \qbar^2 + 4 \pi^2 \Kbar_2 , ~~
%  \beta = - \frac83 \qbar ( 1 - \Kbar_2 ) , ~~
  \beta = - \frac83 ( 1 - \Kbar_2 ) \qbar , ~~
  \gamma = \pi^2 + \Kbar_2 \qbar^2 - \pi^2 \Hbar^2 ,
\end{equation*}
where
\begin{equation}\label{eqn:app_qbaretc}
  \qbar := \frac{\pi}{\vphantom{\overline{L}}\Lbar} , \quad
  \Lbar := \frac{L}{d} , \quad
  \Kbar_2 := \frac{K_2}{K_1} , \quad
  \Hbar := \frac{H}{\Hc} .
\end{equation}
A study of the two eigenvalues of this form leads to a
characterization of the value
\begin{equation*}
  \Kbar_2^{**} := \frac4{3\pi+4} \doteq 0.298 ,
\end{equation*}
such that for $0 < \Kbar_2 < \Kbar_2^{**}$, the approximation
\cref{eqn:app_vw-approx} gives $\delta^2\!\calF[\nhat_0](\bmu) < 0$
for some $A$ and $B$, some $\qbar>0$, and some $\Hbar < 1$.  The
distinguished value emerges in the limit $\qbar \rightarrow 0+$,
$\Hbar \rightarrow 1-$.

Thus the simple approximation \cref{eqn:app_vw-approx} guarantees that
for $0 < K_2 / K_1 < \Kbar_2^{**}$, a periodic instability precedes
the classical \Freed\ transition.  We note that $\Kbar_2^{**}$
compares favorably to the optimal value
\begin{equation}\label{eqn:app_K2barstar}
  \begin{aligned}
    \Kbar_2^* &:= -a + \sqrt{a(a+1)} , ~~ a:= \frac{\,\,\pi^2}8 - 1 \\
    &\phantom{:}\doteq 0.303 ,
  \end{aligned}
\end{equation}
which was found numerically in \cite{lonberg:meyer:85} and by
asymptotics in \cite{oldano:86} and \cite[\S4.3]{virga:94}---below we
give an alternate derivation of $\Kbar_2^*$.  We remark that for
low-molecular-weight liquid crystals (which are typically used in
display applications), one generally finds $K_2 \approx \frac12 K_1$;
whereas for the types of polymer liquid crystals used in
\cite{lonberg:meyer:85}, the authors report much smaller ratios of
$K_2$ to $K_1$, in the range
\begin{equation*}
  10 K_2 < K_1 < 30 K_2 ,
\end{equation*}
which give $\Kbar_2$ well below the value $\Kbar_2^{**}$.

Periodic-in-$y$ modes that result from the substitution of
\cref{eqn:app_vw-series} into \cref{eqn:app_vwEL} are coupled (for the
case of interest $0 < K_2 < K_1$) and satisfy either
\begin{equation}\label{eqn:app_adEVP}
  \begin{gathered}
    K_2 \ahat_k'' + \bigl( \lambda_k - K_1 q_k^2 \bigr) \ahat_k +
    ( K_1 - K_2 ) q_k \dhat_k' = 0 \\
    K_1 \dhat_k'' + \bigl( \chia H^2 + \lambda_k - K_2 q_k^2 \bigr) \dhat_k -
    ( K_1 - K_2 ) q_k \ahat_k' = 0 \\
    \ahat_k(0) = \dhat_k(0) = \ahat_k(d) = \dhat_k(d) = 0 , \quad
    q_k := k \pi / L ,
  \end{gathered}
\end{equation}
with $\bhat_k = \chat_k = 0$, or a similar eigenvalue problem for
$\bhat_k$ and $\chat_k$, with $\ahat_k = \dhat_k = 0$.  Again, the
differential equations uncouple if $K_1 = K_2$.  For an eigenpair
$\ahat_k$, $\dhat_k$, the associated eigenvalue satisfies
\begin{equation*}
  \lambda_k = \frac{\displaystyle \int_0^d \! \Bigl\{
    K_1 \bigl[ (\dhat_k')^2 + q_k^2 \ahat_k^2 \bigr] +
    K_2 \bigl[ (\ahat_k')^2 + q_k^2 \dhat_k^2 \bigr] +
    2 ( K_1 - K_2 ) q_k \ahat_k' \dhat_k -
    \chia H^2 \dhat_k^2 \Bigr\} \, dz}{
    \displaystyle \int_0^d \bigl( \ahat_k^2 + \dhat_k^2 \bigr) \, dz} ,
\end{equation*}
and the local stability is again indicated by the sign of $\lambda_k$.
Solutions of \cref{eqn:app_adEVP} depend only on $L/k$ (not on $L$ and
$k$ independently); so it is sufficient to consider only the case
$k=1$.  We do this and also drop the subscript ``$1$''.

General solutions of the coupled ordinary differential equations in
\cref{eqn:app_adEVP} take different forms depending on $\lambda$.
Three cases can be distinguished (assuming $0 < K_2 < K_1$):
\begin{equation*}
  \mathrm{(I)} ~\, \lambda < K_2 q^2 - \chia H^2 , ~~~~
  \mathrm{(II)} ~\, K_2 q^2 - \chia H^2 < \lambda < K_1 q^2 , ~~~~
  \mathrm{(III)} ~\, K_1 q^2 < \lambda .
\end{equation*}
It can be shown that in Case\,I, there are no nontrivial solutions
that satisfy the boundary conditions.  Case\,III yields an infinite
sequence of \emph{positive} eigenvalues; so it is incapable of
producing an instability.  The relevant case, then, is Case\,II.
Imposing the boundary conditions on the general solution for this case
leads to a transcendental equation that can be solved numerically for
$\lambda$, and this is the approach taken in \cite{lonberg:meyer:85}.
The case is analyzed graphically in \cite[\S4.3]{virga:94}.  Here we
have chosen instead to solve the eigenvalue problem
\cref{eqn:app_adEVP} numerically using a library routine, and for this
we have used the MATLAB\textsuperscript{\textregistered} code
\texttt{bvp5c} \cite{kierzenka:shampine:08}.

In dimensionless terms, the stability eigenvalue problem takes the
form
\begin{equation}\label{eqn:app_abardbarEVP}
  \begin{gathered}
    \Kbar_2 \abar'' + ( \lambdabar - \qbar^2 ) \abar +
    ( 1 - \Kbar_2 ) \qbar \dbar' = 0 \\
    \dbar'' + ( \pi^2 \Hbar^2 + \lambdabar - \Kbar_2 \qbar^2 ) \dbar -
    ( 1 - \Kbar_2 ) \qbar \abar' = 0 \\
    \abar(0) = \dbar(0) = \abar(1) = \dbar(1) = 0 ,
  \end{gathered}
\end{equation}
where
\begin{equation*}
  \zbar := \frac{z}{d} , \quad \abar(\zbar) = \ahat_1(z) , \quad
  \dbar(\zbar) = \dhat_1(z) , \quad \lambdabar := \frac{\lambda_1}{K_1/d^2} ,
\end{equation*}
with $\Kbar_2$, $\qbar$, and $\Hbar$ as previously defined in
\cref{eqn:app_qbaretc}.  For a nontrivial eigenpair $\abar$, $\dbar$,
the associated eigenvalue satisfies
\begin{equation}\label{eqn:app_lambdabar}
  \lambdabar = \frac{\displaystyle \int_0^1 \! \Bigl\{
    \bigl[ (\dbar')^2 + \qbar^2 \abar^2 \bigr] +
    \Kbar_2 \bigl[ (\abar')^2 + \qbar^2 \dbar^2 \bigr] +
    2 ( 1 - \Kbar_2 ) \qbar \abar' \dbar -
    \pi^2 \Hbar^2 \dbar^2 \Bigr\} \, d\zbar}{\displaystyle
    \int_0^1 \bigl( \abar^2 + \dbar^2 \bigr) \, d\zbar} .
\end{equation}
Thus for a given $\Kbar_2$, $\qbar$ (or $\Lbar$), and $\Hbar$, one can
determine (numerically) the mode with the minimal $\lambdabar$ and
adjust $\Hbar$ so that $\lambdabar=0$, giving the critical
magnetic-field strength $\Hbar_\text{p}$ at which the uniform director
field $\nhat = \nhat_0$ becomes unstable with respect to a mode with
that prescribed period:
$\Hbar_\text{p} = \Hbar_\text{p}(\Kbar_2,\qbar)$.

A relevant question is what period gives the earliest instability
onset:
\begin{equation*}
  \Hbar_\text{p}^*(\Kbar_2) =
  \min_{\qbarsub} \Hbar_\text{p}(\Kbar_2,\qbar) .
\end{equation*}
This can be determined as follows.  The dependence of $\lambdabar$ in
\cref{eqn:app_lambdabar} on $\qbar$ is quadratic and can be exhibited
% \begin{equation*}
%   \lambdabar = \frac{\Ibar_0 + \qbar \Ibar_1 + \qbar^2 \Ibar_2}{
%     \displaystyle\int_0^1 \bigl( \abar^2 + \dbar^2 \bigr) \, d\zbar} ,
% \end{equation*}
\begin{equation*}
  \lambdabar = \frac{\Ibar_0 + \Ibar_1 \qbar + \Ibar_2 \qbar^2}{
    \displaystyle\int_0^1 \bigl( \abar^2 + \dbar^2 \bigr) \, d\zbar} ,
\end{equation*}
where
\begin{align*}
  \Ibar_0 &= \int_0^1 \bigl[ (\dbar')^2 + \Kbar_2 (\abar')^2 -
             \pi^2 \Hbar^2 \dbar^2 \bigr] \, d\zbar \\
  \Ibar_1 &= 2 ( 1 - \Kbar_2 ) \int_0^1 \abar' \dbar \, d\zbar \\
  \Ibar_2 &= \int_0^1 \bigl[ \abar^2 + \Kbar_2 \dbar^2 \bigr] \, d\zbar .
\end{align*}
For given, \emph{fixed} functions $\abar$ and $\dbar$, the value of
$\lambdabar$ above will be minimal at
\begin{equation*}
%  \Ibar_1 + 2 \qbar \Ibar_2 = 0 , ~~ \text{if } \Ibar_2 > 0 .
  \Ibar_1 + 2 \Ibar_2 \qbar = 0 , ~~ \text{if } \Ibar_2 > 0 .
\end{equation*}
After a simplification, this gives
\begin{equation}\label{eqn:app_qstar}
  \qbar^* = \frac{\pi}{\,\Lbar^*\vphantom{\overline{L}}} =
  \frac{\displaystyle(1-\Kbar_2) \int_0^1 \abar\dbar' \, d\zbar}{
    \displaystyle\int_0^1 \bigl( \abar^2 + \Kbar_2 \dbar^2 \bigr) \, d\zbar} .
\end{equation}

Thus, to obtain the instability mode with the optimal period (and
smallest required magnetic-field strength), one must solve the
stability eigenvalue problem \cref{eqn:app_abardbarEVP} with $\qbar$
(or $\Lbar$) subject to the integral constraint \cref{eqn:app_qstar}.
We have done this by a simple decoupling iteration (as a matter of
expediency): solving \cref{eqn:app_abardbarEVP} with a given $\qbar$,
computing the ``optimal'' $\qbar$ associated with that solution using
\cref{eqn:app_qstar}, re-solving \cref{eqn:app_abardbarEVP} with this
new $\qbar$, etc., iterating until convergence.  The results for some
representative values are given in \cref{tab:app_Hpstars}.
\begin{table}\label{tab:app_Hpstars}
  \caption{Minimal reduced magnetic-field strength $\Hbar_\text{p}^*$
    of periodic instability and associated optimal half period
    $\Lbar^*$ (in units of the cell gap) as a function of the ratio of
    the twist elastic constant $K_2$ to the splay elastic constant
    $K_1$ for some representative values: $\Kbar_2=K_2/K_1$,
    $\Hbar=H/\Hc$ (with $\Hc$ as defined in \cref{eqn:app_Hc}),
    $\Lbar=L/d$.}
  \centering
  \begin{tabular}{|c|c|c|} \hline
    $\Kbar_2\vphantom{\hat{K}}$ & $\Hbar_\text{p}^*$  & $\Lbar^*$ \\ \hline
    0.10 & 0.753 & 0.822 \\
    0.15 & 0.871 & 0.955 \\
    0.20 & 0.945 & 1.175 \\
    0.25 & 0.986 & 1.652 \\ \hline
  \end{tabular}
\end{table}
We note that in the experiments reported in \cite{lonberg:meyer:85}, a
period of 65\,$\mu$m was observed for a fully developed periodic
solution for a material with $\Kbar_2 < 0.10$ in a cell of thickness
37\,$\mu$m, which corresponds to $\Lbar \doteq 0.878$---the
periodicities reported in \cref{tab:app_Hpstars} are at the onset of
the instability.

The period of the instability at onset diverges as $\Kbar_2$
approaches the limiting value $\Kbar_2^*$:
\begin{equation*}
  \Kbar_2 \rightarrow \Kbar_2^* ~ \Rightarrow ~
  \qbar^* \rightarrow 0 , ~ \Lbar^* \rightarrow \infty .
\end{equation*}
For $\Kbar_2=0.3$ (which is within 1\% of $\Kbar_2^*$), our numerics
give $\Hbar_\text{p}^* \doteq 0.99995$, $\Lbar^*\doteq 6.75$.  The
vanishing of $\qbar$ in this limit is what enabled Oldano in
\cite{oldano:86} to determine the analytical formula
\cref{eqn:app_K2barstar} for $\Kbar_2^*$.  The approach taken in
\cite{oldano:86} (also used in \cite[\S4.3]{virga:94}) was to set
$\Hbar=1$ and $\lambdabar=0$ in the transcendental equation that
results from imposing the homogeneous boundary conditions on the
general solution of the differential equations in
\cref{eqn:app_abardbarEVP}, expand in powers of $\qbar$ (out to
$O(\qbar^2)$), simplify, and solve for $\Kbar_2$ in the limit
$\qbar\rightarrow0$.  Here we show, in a similar vein, how $\Kbar_2^*$
can be obtained from a perturbation expansion in the stability
eigenvalue problem.

We work from the problem in dimensionless form
\cref{eqn:app_abardbarEVP}, subject to the convenient normalization
\begin{equation*}
  \dbar'(0)=\pi
\end{equation*}
and the optimal-$\qbar$ integral constraint \cref{eqn:app_qstar},
which we write
\begin{equation*}
  (1-\Kbar_2) \int_0^1 \abar' \dbar \, d\zbar +
  \qbar \int_0^1 \bigl( \abar^2 + \Kbar_2 \dbar^2 \bigr) \, d\zbar = 0 .
\end{equation*}
We use $\qbar$ as the expansion parameter (since the solution of
interest emerges with $\qbar=0$).  Dropping bars, we substitute the
formal expansions
% \begin{gather*}
%   a = a_0 + q a_1 + q^2 a_2 + \cdots , \quad
%   d = d_0 + q d_1 + q^2 d_2 + \cdots , \\
%   \lambda = \lambda_0 + q \lambda_1 + q^2 \lambda_2 + \cdots , \quad
%   H = H_0 + q H_1 + q^2 H_2 + \cdots
% \end{gather*}
\begin{gather*}
  a = a_0 + a_1 q + a_2 q^2 + \cdots , \quad
  d = d_0 + d_1 q + d_2 q^2 + \cdots , \\
  \lambda = \lambda_0 + \lambda_1 q + \lambda_2 q^2 + \cdots , \quad
  H = H_0 + H_1 q + H_2 q^2 + \cdots
\end{gather*}
into the differential equations, boundary conditions, normalization
condition, and integral constraint.  At order $O(1)$, we obtain
\begin{gather*}
  K_2 a_0'' + \lambda_0 a_0 = 0 , ~~ a_0(0) = a_0(1) = 0 , \\
  d_0'' + ( \pi^2 H_0^2 + \lambda_0 ) d_0 = 0 , ~~
  d_0(0) = d_0(1) = 0 , ~ d_0'(0) = \pi .
\end{gather*}
At the point of interest, we have $\lambda_0 = 0$ (the threshold of
the periodic instability), which implies $a_0=0$ and leads to a family
of solutions for $d_0$ with $H_0 = \pm1, \pm2, \ldots ,$ the one of
interest being $H_0=1$:
\begin{equation*}
  a_0 = 0 , \quad d_0 = \sin \pi z .
\end{equation*}

At order $O(q)$, we have
\begin{gather*}
  K_2 a_1'' + ( 1 - K_2 ) d_0' = 0 , ~~ a_1(0) = a_1(1) = 0 , \\
  d_1'' + \pi^2 d_1 + ( 2 \pi^2 H_1 + \lambda_1 ) d_0 = 0 , ~~
  d_1(0) = d_1(1) = d_1'(0) = 0 , \\
  ( 1 - K_2 ) \int_0^1 a_1' d_0 \, dz + K_2 \int_0^1 d_0^2 \, dz = 0 .
\end{gather*}
The $a_1$ solution is given by
\begin{equation*}
  a_1 = \frac{1-K_2}{K_2\pi} ( \cos \pi z + 2 z - 1 ) ,
\end{equation*}
while the solvability condition for the differential equation for
$d_1$ requires
\begin{equation*}
  2 \pi^2 H_1 + \lambda_1 = 0 ,
\end{equation*}
leaving
\begin{equation*}
  d_1 = 0 .
\end{equation*}
The integral constraint gives
\begin{equation*}
  ( 1 - K_2 ) \int_0^1 a_1' d_0 \, dz + K_2 \int_0^1 d_0^2 \, dz =
  \frac{(1-K_2)^2}{K_2\pi} \Bigl( \frac4{\pi} - \frac{\pi}2 \Bigr) +
  \frac{K_2}2 = 0 ,
\end{equation*}
which simplifies to
\begin{equation*}
  K_2^2 + 2 a K_2 - a = 0 , \quad a := \frac{\,\,\pi^2}8 - 1 ,
\end{equation*}
for which the positive root is
\begin{equation*}
  K_2 = - a + \sqrt{ a ( a + 1 ) } .
\end{equation*}
This is precisely the quadratic polynomial and root formula for
$\Kbar_2^*$ given in \cite{oldano:86} and \cite[Th.\,4.9]{virga:94}
(modulo a typographical sign error in \cite[(4.85)]{virga:94}).  We
remark that the reason this technique works here is that $q_0=0$, and
the differential equations uncouple at leading order.  The approach
does not lead to simple analytical solutions when $\Kbar_2<\Kbar_2^*$,
where $q_0\not=0$ at the bifurcation point and the equations for $a_0$
and $d_0$ remain coupled (and require numerical methods at some
stage).

\section{Inclusion of flexoelectric effects}
\label{app:flexoelectric}
The models considered thus far have been deliberately kept as simple
as possible so as to focus on the coupling between the director field
and the electric field and its consequences with respect to local
stability of equilibrium solutions.  The approach and ideas, however,
are general, and here we provide an illustration of how an additional
feature can be incorporated into the theory: ``flexoelectricity.''
Flexoelectricity concerns polarization caused by director distortion,
and flexoelectric effects sometimes play an important role in liquid
crystal systems---see \cite[\S3.3.2]{degennes:prost:93} or
\cite[\S4.1]{lagerwall:99}.  Since these effects involve an interplay
between director distortions and electric fields, it is natural to
wonder about how they would fit into our development.

We use the same building blocks that we have used previously and
consider a free-energy functional of the form
\begin{equation*}
  \calF[\nhat,\varphi] = \int_\Omega
  W(\nhat,\nabla\nhat,\nabla\varphi) \, dV +
  \int_{\Gamma_2} \Ws(\nhat) \, dS ,
\end{equation*}
with $\Omega$ and $\Gamma_2$ as depicted in \cref{fig:domain} and with
$\Ws$ an appropriate surface anchoring energy, as in
\cref{sec:model-problems}.  However, the free-energy density now is
given by
\begin{gather*}
  W = \We(\nhat,\nabla\nhat) -
  \frac12 \epstensor(\nhat) \nabla\varphi \cdot \nabla\varphi +
  \Pf(\nhat,\nabla\nhat) \cdot \nabla\varphi , \\
  \Pf = \es ( \div\nhat ) \nhat + \eb \nhat \times \curl\nhat .
\end{gather*}
% \begin{equation*}
%   W = \We(\nhat,\nabla\nhat) -
%   \frac12 \epstensor(\nhat) \nabla\varphi \cdot \nabla\varphi +
%   \Pf(\nhat,\nabla\nhat) \cdot \nabla\varphi , ~~
%   \Pf = \es ( \div\nhat ) \nhat + \eb \nhat \times \curl\nhat .
% \end{equation*}
The first two terms of $W$ here are as before: $\We$, the distortional
elasticity (as in \cref{eqn:We}), and the dielectric tensor
${\mathlarger\epstensor}$ as in \cref{eqn:WE}.  The third term is the
new addition, with $\Pf$ denoting the flexoelectric polarization and
$\es$ and $\eb$ the ``splay'' and ``bend'' flexoelectric coefficients
(which can be positive or negative).  This term accounts for the
phenomenon of splay deformations and bend deformations inducing
polarization.  We note that the flexoelectric term is linear in
$\nabla\varphi$ (whereas the second term of $W$ above is quadratic)
and that it also introduces a coupling between $\bmE$ and
$\nabla\nhat$ (in addition to the coupling between $\bmE$ and $\nhat$
already present in the second term).

The main results of \cref{sec:stability-for-E-fields} remain valid.
Here we highlight the changes caused by the addition of the
flexoelectric term.  The problem that determines the electric
potential $\varphi$ from a given director field $\nhat$
($\varphi=T(\nhat)$, $\delta_\varphi\calF[\nhat,\varphi]=0$) now reads
\begin{subequations}\label{eqn:app_phi-strong-flexo}
\begin{equation}
  \int_\Omega \epstensor(\nhat) \nabla\varphi \cdot \nabla\psi \, dV =
  \int_\Omega \Pf(\nhat,\nabla\nhat) \cdot \nabla\psi \, dV , ~~
  \forall \psi \in \PSI_0
\end{equation}
or
\begin{equation}
  \begin{gathered}
    \div \bigl[ \epstensor(\nhat) \nabla\varphi \bigr] =
    \div \bigl[ \Pf(\nhat,\nabla\nhat) \bigr] \text{ in } \Omega \\
    \varphi = 0 \text{ on } \Gamma_1 , ~
    \varphi = V \text{ on } \Gamma_2 , ~
    \varphi \text{ periodic on } \Gamma_3 .
  \end{gathered}
\end{equation}
\end{subequations}
The equilibrium equations for the director field have the same form as
before:
\begin{gather*}
  - \div \Bigl( \dWdgn \Bigr) + \dWdn =
  \lambda \nhat \text{ in } \Omega , \\
  \nhat = \nhatb \text{ on } \Gamma_1 , ~~
  \Bigl( \dWdgn \Bigr) \nuhat + \dWsdn =
  \mu \nhat \text{ on } \Gamma_2 , ~~
  \nhat \text{ periodic on } \Gamma_3 .
\end{gather*}
There are, however, new contributions to both $\partial W / \partial
\nabla\nhat$ and $\partial W / \partial \nhat$ from the term
$\Pf(\nhat,\nabla\nhat) \cdot \nabla\varphi$, which we do not expand
upon here---see \cite[\S5.3]{gartland:20}.

A perturbation of an equilibrium director field
($\nhat_0 \mapsto \nhat_0 + \eps \bmv$) will cause changes in both the
electric-field-induced polarization (at first order, $\bmd_0$, as
before in \cref{eqn:dzero})
\begin{equation*}
    \bmd_0 = \epsz \epsa (
    \nhat_0\otimes\bmv + \bmv\otimes\nhat_0 ) \bmE_0 , \quad
    \bmE_0 = - \nabla\varphi_0
\end{equation*}
and also now in the director-distortion-induced polarization, which at
first order is given by
\begin{equation*}
  \bmP_0 = \es \bigl[ (\div\nhat_0)\bmv + (\div\bmv)\nhat_0 \bigr] +
  \eb ( \nhat_0\times\curl\bmv + \bmv\times\curl\nhat_0 ) .
\end{equation*}
If the ground-state director field is uniform
($\nhat_0=\text{const}$), then this simplifies to
\begin{equation*}
  \bmP_0 = \es (\div\bmv)\nhat_0 + \eb \nhat_0\times\curl\bmv .
\end{equation*}
Thus the problem of determining the first-order change in the electric
potential ($\psi = DT(\nhat_0) \bmv$) now takes the form
\begin{subequations}\label{eqn:app_psi-strong-flexo}
\begin{equation}
  \int_\Omega \epstensor(\nhat_0) \nabla\psi \cdot \nabla\chi \, dV =
  \int_\Omega ( \bmd_0 + \bmP_0 ) \cdot \nabla\chi \, dV , ~~
  \forall \chi \in \PSI_0
\end{equation}
or
\begin{equation}
  \begin{gathered}
    \div \bigl[ \epstensor(\nhat_0) \nabla\psi \bigr] =
    \div ( \bmd_0 + \bmP_0 ) \text{ in } \Omega \\
    \psi = 0 \text{ on } \Gamma_1 \text{ and } \Gamma_2 , ~~
    \psi \text{ periodic on } \Gamma_3 .
  \end{gathered}
\end{equation}
\end{subequations}
Thus $\psi = 0$ on $\Omega$ if and only if $\bmd_0 + \bmP_0$ is
divergence free on $\Omega$, and we now have
\begin{equation*}
  \int_\Omega ( \bmd_0 + \bmP_0 ) \cdot \nabla\psi \, dV =
  \int_\Omega \epstensor(\nhat_0) \nabla\psi \cdot \nabla\psi \, dV .
\end{equation*}

The second-order necessary condition for the local stability of the
equilibrium pair $\nhat_0$, $\varphi_0$ reads exactly as before in
\cref{eqn:second-order-electric}:
\begin{multline*}
  \delta_{\nhat\nhat}^2 \calF[\nhat_0,\varphi_0](\bmu) +
  \int_\Omega \epstensor(\nhat_0) \nabla\psi \cdot \nabla\psi \, dV \\
  {} - \int_\Omega \lambda_0 |\bmu|^2 \, dV -
  \int_{\Gamma_2} \mu_0 |\bmu|^2 \, dS \ge 0 , ~~
  \forall \bmu \in \calU_0 .
\end{multline*}
Here, however, $\varphi_0$ and $\psi$ satisfy the slightly modified
problems \cref{eqn:app_phi-strong-flexo} and
\cref{eqn:app_psi-strong-flexo}.  Our previous conclusions and
interpretations remain valid.  As before, everything hinges on whether
or not perturbations of the equilibrium director field
($\nhat_0 \mapsto \nhat_0 + \eps \bmu$, $\bmu \in \calU_0$) cause a
first-order change in the electric field
($\bmE = \bmE_0 + \delta\bmE$).  If $\div (\bmd_0+\bmP_0) = 0$ on
$\Omega$, then $\psi=0$, and $\delta\bmE = \bfzero$, and the
$\nhat$-$\bmE$ coupling has no effect on the instability threshold.
Otherwise the coupling will elevate the threshold.  It is the case now
that it is the combined effect of the first-order changes in the
electric-field-induced polarization ($\bmd_0$) and the
director-induced-polarization ($\bmP_0$) that determines the outcome.

A natural question at this point is whether or not the flexoelectric
terms (through $\bmP_0$ and $\div\bmP_0$) could have any effect on
instability thresholds such as \Freed\ transitions.  We show now that
this is not the case, at least for \Freed\ transitions: none of the
classical \Freed\ transitions are altered by the inclusion of $\Pf$ in
the free-energy density.  First, note that flexoelectric effects can
play a role in liquid crystal systems with magnetic fields, as well as
those with electric fields.  The role of flexoelectricity in the
magnetic-field splay-\Freed\ transition is analyzed in
\cite{deuling:74}; while the electric-field splay-\Freed\ transition
(with flexoelectric terms included) is studied via experiment and
theory in \cite{brown:mottram:03}.  In both cases, flexoelectric
effects were explored above the instability threshold, while the
threshold itself was found not to be affected by the inclusion of the
flexoelectric terms.  Next, recall that for all of the classical
\Freed\ transitions, the ground-state director field $\nhat_0$ is
uniform.  Thus $\nabla\nhat_0 = \bfzero$, and in this case, $\bmP_0$
is given by
\begin{equation*}
  \bmP_0 = \es ( \div\bmu ) \nhat_0 + \eb \nhat_0\times\curl\bmu .
\end{equation*}
It is also the case that $\bmP_0$ is independent of the electric field
(by virtue of the fact that the coupling $\Pf\cdot\bmE$ is linear in
$\bmE$).  Thus $\bmP_0$ depends only on $\nhat_0$ and $\nabla\bmu$,
which leaves us with just three geometries and symmetry assumptions to
consider.

In the splay geometry (depicted in \cref{fig:Freed-geoms-left}),
\begin{equation*}
  \nhat_0 = \ehat_x , ~ \bmu = w(z) \ehat_z ~ \Rightarrow ~
  \div \bmu = w_z , ~ \curl \bmu = \bfzero ,
\end{equation*}
which gives
\begin{equation*}
  \bmP_0 = \es w_z \ehat_x ~ \Rightarrow ~ \div \bmP_0 = 0 .
\end{equation*}
Thus the coupling between $\Pf$ and $\bmE$ in this case can have no
effect on the instability threshold, no matter if the instability is
induced by a magnetic field or an electric field or if the magnetic or
electric anisotropy is positive or negative.  The twist geometry, as
usually written in the textbooks (which we have not depicted),
corresponds to
\begin{equation*}
  \nhat_0 = \ehat_y , ~ \bmu = u(z) \ehat_x ~ \Rightarrow ~
  \div \bmu = 0 , ~ \curl \bmu = u_z \ehat_y ,
\end{equation*}
using the same coordinate system as in \cref{fig:Freed-geoms}.  This
gives
\begin{equation*}
  \nhat_0 \times \curl \bmu = 0 ~ \Rightarrow ~
  \bmP_0 = \bfzero ~ \Rightarrow ~ \div \bmP_0 = 0 ,
\end{equation*}
and the coupling does not affect the threshold again in this case.
Finally, the bend geometry (depicted in \cref{fig:Freed-geoms-center})
corresponds to
\begin{equation*}
  \nhat_0 = \ehat_z , ~ \bmu = u(z) \ehat_x ~ \Rightarrow ~
  \div \bmu = 0 , ~ \curl \bmu = u_z \ehat_y ,
\end{equation*}
which gives
\begin{equation*}
  \bmP_0 = - \eb u_z \ehat_x ~ \Rightarrow ~ \div \bmP_0 = 0 ,
\end{equation*}
and the coupling is again ineffectual.  These results are consistent
with \cite{brown:mottram:03,deuling:74} in the case of splay
transitions.  To our knowledge, these observations are new for the
twist and bend geometries (though consistent with what most assume to
be true).

\section*{Acknowledgments}

The author is grateful to P.~Palffy-Muhoray for useful information and
references on \Freed\ transitions and experiments with in-plane fields
and also to O.~D.~Lavrentovich for helpful feedback on a presentation
of a preliminary version of this report.

%\nocite{*}

\bibliography{paper}

\end{document}